\documentclass[11pt,a4paper]{article}

\usepackage{jheppub}
\bibstyle{JHEP}
\usepackage[english]{babel}
\pdfoutput=1
\usepackage{physics}
\usepackage{tikz}
\usepackage{jheppub}
\usepackage{amsmath,amssymb,amsfonts}
\usepackage{graphicx}
\usepackage{color}
\usepackage{tikz}
\usetikzlibrary{decorations.markings,patterns}
\usepackage{dsfont}
\usepackage{subfigure}
\usepackage{xcolor}
\usepackage{hyperref}
\usepackage[colorlinks=true,urlcolor=blue,linkcolor=blue,citecolor=blue,linktocpage=true]{hyperref}
\usetikzlibrary{calc}   
\definecolor{darkred}{rgb}{0.8,0.1,0.1}
\hypersetup{colorlinks=true, linkcolor=darkred, citecolor=blue, linktoc=page}

\definecolor{3dcolor}{rgb}{0.96,0.89,0.76}
\definecolor{4dcolor}{rgb}{0.812,0.851,0.914}

\def\CC{{\mathds{C}}}
\def\RR{{\mathds{R}}}

\def\cY{{\cal Y}}

\newcommand{\be}{\begin{equation}}
	\newcommand{\ee}{\end{equation}}
\newcommand{\ba}{\begin{array}}
	\newcommand{\ea}{\end{array}}
\newcommand{\bea}{\begin{equation}\begin{aligned}}
		\newcommand{\eea}{\end{aligned}\end{equation}}

\newcommand{\mo}{\mathcal{O}}

\newcommand{\p}{\partial}

\newcommand{\mrr}{\mathbb{R}}

\newcommand{\pderi}[2]{\frac{\p #1}{\p #2}}

\newcommand{\ads}{\text{AdS}}

\newcommand{\mn}{\mu\nu}

\newcommand{\lbra}[1]{\left( #1 \right)}
\newcommand{\lmid}[1]{\left[ #1 \right]  }
\newcommand{\thes}{\theta_{\star}}

\newcommand{\ap}{\alpha'}
\newcommand{\wo}{W_{\mathrm{out}}}
\newcommand{\wi}{W_{\mathrm{in}}}
\newcommand{\xp}{x_{\perp}}
\newcommand{\con}{\frac{\mu_0^2}{\kappa^2}}
\newcommand{\er}{e_{\mrr}}

\title{Resolving BCFT bulk-cone singularities in braneworlds and top-down holography}

\author[a]{Dongming He}
\author[a]{and Christoph F.~Uhlemann}

\affiliation[a]{Theoretische Natuurkunde, Vrije Universiteit Brussel and The International Solvay Institutes, Pleinlaan 2, B-1050 Brussels, Belgium}

\emailAdd{dongming.he@vub.be}
\emailAdd{christoph.uhlemann@vub.be}

\abstract{Two-point functions in BCFTs exhibit characteristic singularities compared to standard CFTs due to reflected light cones. In holographic BCFTs, further singularities, which need not be related to reflected light cones within the BCFT, can arise when the insertion points are connected by bulk null geodesics. The resulting `bulk-cone singularities' are expected to be an artifact of the `t Hooft limit and should be resolved in string theory. In this work we demonstrate the resolution of bulk-cone singularities in braneworld BCFT duals, where bulk-cone singularities arise from geodesics reflected off the end-of-the-world brane. We show that the geometry near the null geodesics is described by a pp-wave of shockwave form and incroporate string theory corrections, which bounds the two-point functions. We extend this discussion to top-down holography for D1/D5 BCFTs.	
}

\begin{document}
	\maketitle

\section{Introduction}
	
Singularities of correlation functions encode fundamental features of quantum field theories, such as the presence of on-shell states and the singular parts of operator product expansions.
	
In conformal field theories (CFTs), correlation functions computed in perturbation theory should only have singularities at points where null geodesics emerging from each insertion point intersect at a common point with momentum locally conserved. This is captured by Landau diagrams  \cite{Maldacena:2015iua}. In CFTs with gravity duals there can be additional singularities in the semi-classical limit. These should only be approximate. For example, bulk-point singularities arise when the cross-ratios $\xi$ of four-point functions satisfy $\xi=\bar{\xi}$ \cite{Gary:2009ae}, and have to be resolved at the scale where the gravity theory becomes non-local \cite{Maldacena:2015iua}. Similar approximate singularities exist in CFTs at finite temperature whose gravity duals are black holes \cite{Hubeny:2006yu,Dodelson:2023nnr}. Two-point functions with insertion points connected by bulk null geodesics display bulk-cone singularities that are resolved by stringy effects \cite{Dodelson:2020lal}.

Such approximate singularities also arise in holographic CFTs with boundaries. In bottom-up braneworld models, holographic duals of BCFTs are modeled as asymptotically-AdS spacetime terminated by an end-of-the-world (ETW) brane \cite{Karch:2000ct,Karch:2000gx,Takayanagi:2011zk,Fujita:2011fp}. Two-point functions with insertion points connected by null geodesics scattered off the ETW brane in the bulk also exhibit bulk-cone singularities \cite{Reeves:2021sab,Kastikainen:2021ybu}\footnote{Such singularities were called bulk-brane singularities in \cite{Reeves:2021sab}. Since ETW-branes are resolved into geometry in top-down constructions, we still call them bulk-cone singularities in this work.\label{footnote-bulk-brane}}. As argued in \cite{Maldacena:2015iua, Reeves:2021sab}, such singularities should also be approximate and should be resolved in string theory. In top-down string theory constructions, the ETW branes are replaced by higher-dimensional internal spaces where the geometry terminates smoothly by collapsing internal cycles. In 4d, a rich class of BCFTs can be engineered by D3/D5/NS5 systems \cite{Gaiotto:2008sa, Gaiotto:2008ak}, with holographic duals with $\ads_4\times S^2 \times S^2 \times \Sigma$ geometry in type IIB string theory \cite{DHoker:2007zhm,DHoker:2007hhe,Aharony:2011yc,Assel:2011xz} (recent developments include \cite{He:2024djr,He:2025etu}). In 2d, there are BCFTs engineered by D1/D5 systems whose holographic duals can be constructed in truncations to 6d gauged supergravity as solutions with $\ads_2\times S^2 \times \Sigma$ geometry  \cite{Chiodaroli:2011nr,Chiodaroli:2011fn,Chiodaroli:2012vc}. Most null geodesics starting at the conformal boundary representing the ambient space in the BCFT get trapped in the internal space, as we will discuss, but certain fine-tuned geodesics can still lead to bulk-cone singularities \cite{Reeves:2021sab,Karch:2022rvr}. The question for the resolution of these singularities therefore arises in bottom-up and top-down models.

In this work we study the resolution of bulk-cone singularities in holographic BCFTs by stringy $\ap$ corrections. When the insertion points are connected by bulk null geodesics, the behavior of a two-point function is dominated by the geometry near that geodesic. The $\ap$ corrections from this region in turn become tractable, since the geometry near null geodesics simplifies. It is determined by the Penrose limit \cite{Penrose:1976pw} and generically leads to a pp-wave with metric
\bea\label{eq:pp-wave-0}
	ds^2=2dUdV+A_{\alpha\beta}(U) x^{\alpha} x^{\beta} dU^2+ d\vec{x}^2~,
\eea
where $U,V$ are null coordinates and $x^{\alpha}$ are transverse coordinates. The string action becomes quadratic on this background and quantization becomes tractable \cite{Metsaev:2001bj,Blau:2024}. The world-sheet equations of motion become linear with masses set by the pp-wave profile $A_{\alpha\beta}(\tau)$. The Penrose limit captures tidal forces acting on strings, leading to particle production effects on the worldsheet. For geodesics in $\ads_5\times S^5$ with angular momentum on $S^5$, the Penrose limit leads to constant masses and describes BMN sectors of $\mathcal{N}=4$ SYM \cite{Berenstein:2002jq}, and for BMN-like sectors in BCFTs see \cite{Chaney:2024bgx}. In general Penrose limits the pp-wave profiles $A_{\alpha\beta}$ are time-dependent (e.g.~\cite{Martinec:2020cml}), and the particle production effects can be understood similar to those of quantum field theory in curved spacetime, as e.g.\ in \cite{Dodelson:2020lal}.
	
Our first focus will be on braneworld models, whose study is motivated e.g.\ by their applications in studying Euclidean wormholes and the black hole information problem. Braneworld models are bottom-up constructions and not consistent string theory backgrounds, but we will proceed as if at least the pp-wave geometry obtained as Penrose limit were part of a consistent 10d string theory background, and study strings propagating on this geometry. This is a relatively mild assumption. We will comment on the connection to top-down holographic duals for BCFTs in the main part and justify this assumption.
	
A useful fact is that Penrose limits of pure AdS spacetimes are actually flat \cite{Blau:2024,Siopsis:2002vw}, with vanishing pp-wave profile $A_{\alpha\beta}$. In the simplest braneworld models, with undeformed  AdS cut off by an ETW brane, the Penrose limit therefore takes a shockwave form 
\bea \label{eq: metric-shock-pp-pre}
	ds^2=2dU dV+\delta(U) A_{\alpha\beta}x^{\alpha}x^{\beta}\,dU^2+d\vec{x}^2~,
\eea
with the shock induced by the ETW brane. We will derive $A_{\alpha\beta}$, which depends on the brane tension, and show how the bulk-cone singularities are resolved. In general, the asymptotically-AdS bulk that is terminated by the ETW brane should not be expected to be undeformed AdS even in braneworld models.\footnote{One reason is that BCFTs typically exhibit non-trivial one-point functions, which are allowed by the symmetries. This leads to non-trivial profiles for the dual bulk fields whose backreaction deforms AdS.\label{foot:ads}} Nevertheless, string fluctuations with large light-cone momentum still experience a shockwave geometry, and we will extend the discussion to braneworld models with general asymptotically-$\ads_{d+1}$ geometries terminated by ETW branes as well as to actual top-down holographic duals for D1/D5 BCFTs.
	
\smallskip
	
\textbf{Outline:}
In section \ref{sec: review-sin} we review kinematic singularities in BCFTs and bulk-cone singularities in holographic BCFTs. We discuss the geodesics leading to bulk-cone singularities in braneworld models and the Penrose limit in section \ref{sec: Pen-brane}, and demonstrate the resolution of bulk-cone singularities by stringy effects in section \ref{sec: resolve}. In section \ref{sec: D1-D5} we extend the discussion to top-down holographic duals for D1/D5 BCFTs. We conclude in section \ref{sec: dis}.

\section{Review of singularities in BCFT correlation functions}\label{sec: review-sin}
Following \cite{Reeves:2021sab}, we review singularities of two-point functions in BCFTs.
We first review general kinematic singularities. Then we show that in bottom-up models with simple holographic duals, a new type of singularities named bulk-cone singularities (or bulk-brane singularities as in footnote~\ref{footnote-bulk-brane}) can appear when the insertion points are connected by a bulk geodesic scattered off the ETW brane. 
Related singularities in top-down BCFTs are discussed in section \ref{sec: D1-D5}.
Readers familiar with the singularities can skip this section.

\subsection{Kinematic singularities in general BCFTs}
Singularities in BCFT two-point functions should be kinematic and independent of the theory. Consider a $d$-dimensional BCFT on a half-plane $  \mrr^{d-1} \times \mrr^+$ with coordinates $x=((x_0,\vec{x})\in\mrr^{d-1} , x_{\perp} \in \mrr^+ )$.
Due to the reduced symmetries, one-point functions of ambient operators are in general non-vanishing.
Similar to two-point functions in CFTs, their form is fixed up to a coefficient $a_{\mo}$, to take the form
\bea
\expval{\mathcal O(x)}=\frac{a_{\mo}}{\xp^{\Delta_{\mo}}}~.
\eea
With the operator normalization fixed independently, $a_{\mo}$ contains dynamical information.
Two-point functions in BCFTs are structurally similar to four-point functions in CFTs and less constrained in their form.
From two points one can form a conformally invariant cross ratio $\xi$ and the form of the two-point function is fixed up to a function of this cross ratio,
\begin{align}\label{eq: cross-ratio}
	\langle\mathcal{O}(x) \mathcal{O}(y)\rangle&=\frac{\mathcal{G}(\xi)}{\left|4 x_{\perp} y_{\perp}\right|^{\Delta}}~,
	&
	\xi&=\frac{(x-y)^2}{4 x_{\perp} y_{\perp}}~.
\end{align}

In Lorentzian CFTs, two-point functions have universal singularities when one insertion point approaches the light cone of the other, i.e.\ $(x-y)^2\to 0$ or $\xi=0$.
This also holds in BCFTs.
In BCFTs, light cones can further be reflected at the boundary, and a new type of singularity appears when one insertion point approaches the reflected light cone of the other. This corresponds to $\xi=-1$ \cite{Mazac:2018biw}. It has been argued that these two are the only true singularities in BCFTs \cite{Reeves:2021sab, Maldacena:2015iua}.

\subsection{Bulk-cone singularities in bottom-up BCFT holography}\label{sec: bulk-cone BCFT}

In this section we review how bulk-cone singularities arise in holographic duals of BCFTs. We start with a particularly simple braneworld model and review the relevant geodesics, and then comment on more general braneworld models and top-down holographic duals.

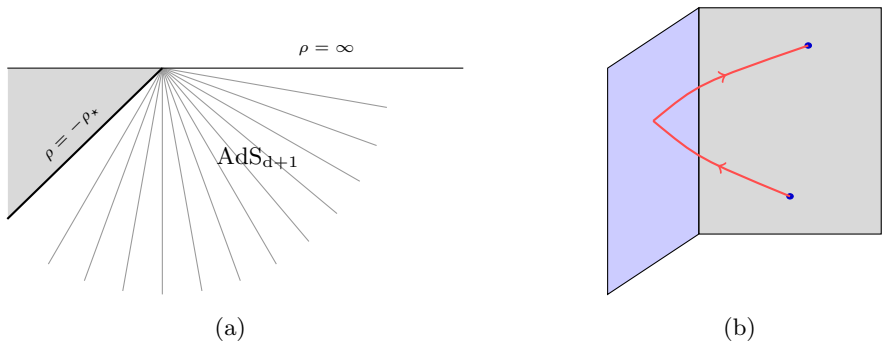
\begin{figure}
	\centering
	\subfigure[][]{\label{fig:AdS-brane}
		\begin{tikzpicture}[scale=1.2]
			\foreach \i in {-10,-20,...,-120}{ \draw[black!40] (-0.8,0) -- +({2.5*cos(\i)},{2.5*sin(\i)});};
			
			\draw [white,fill=gray,opacity=0.3] (-0.8,0) -- (-2.5,0) -- (-2.5,-2/3*2.5)--(-0.8,0);
			
			\draw (-2.5,0) -- (2.5,0);		
			\draw[thick] (-0.8,0) -- (-2.5,-2/3*2.5);
			
			\node [rotate=45] at (-1.8,-0.7) {\tiny $\rho=-\rho_\star$};
			\node at (1,0.2) {\tiny $\rho=\infty$};
			\node at (0.25,-1) {\footnotesize $\rm AdS_{d+1}$};
		\end{tikzpicture}
	}
	\hskip 15mm
	\subfigure[][]{\label{fig:Empty}
		\begin{tikzpicture}[xscale=1.2,yscale=1]
			\fill[gray!30] (0,0) -- (0,3) -- (2,3) -- (2,0) -- (0,0);
			
			\fill[blue!20] (0,0) -- (-1,-0.8) -- (-1,3-0.8) -- (0,3) -- (0,0);
			\draw[black] (0,0) -- (0,3) -- (2,3) -- (2,0) -- (0,0);
			\draw[black] (0,0) -- (-1,-0.8) -- (-1,3-0.8) -- (0,3) -- (0,0);
			\filldraw[blue!80!black] (1,0.5) circle (1pt);
			\filldraw[blue!80!black] (1.2,2.5) circle (1pt);
			
			\draw[red!70, thick, decoration={
				markings,
				mark=at position 0.5 with {\arrow{>}}
			},
			postaction={decorate}]
			(1,0.5) .. controls (0,1)  .. (-0.5,1.5);
			\draw[red!70, thick, decoration={
				markings,
				mark=at position 0.5 with {\arrow{>}}
			},
			postaction={decorate}]
			(-0.5,1.5) .. controls (0,2)  .. (1.2,2.5);
			
		\end{tikzpicture}
	}
	\caption{Left: Braneworld AdS/BCFT geometry with ETW brane at $\rho=\rho_{\star}$. $\ads_d$ slices are drawn as thin gray lines. The asymptotic boundary is at $\rho=\infty$. Right: Null geodesics (red) emerging from the boundary (grey) and reflected by the brane (blue).}
\end{figure}

For the bottom-up discussion we consider a simple model where the bulk manifold $\mathcal{M}$ is terminated by an ETW brane $\mathcal{B}$, which acts as an additional boundary and meets with the asymptotic boundary of $\mathcal{M}$ at the BCFT boundary.
The action is
\begin{equation} \label{eq: action-brane}
	S_G=\frac{1}{16 \pi G_N} \int_{\mathcal{M}} \sqrt{-g}(R-2 \Lambda)+\frac{1}{8 \pi G_N} \int_{\mathcal{B}} \sqrt{-h}(K-T)~.
\end{equation}
The variational principle with Neumann boundary condition on the brane yields
\bea \label{eq: junction}
K_{ab}=(K-T)h_{ab}~,
\eea
which determines the location of the brane in terms of the tension $T$.
A particular bulk solution is $\ads_{d+1}$, whose metric can be expressed in terms of $\ads_d$ slices as (fig.\ \ref{fig:AdS-brane})
\bea \label{eq: ads metric slice}
ds^2=L^2(d\rho^2+\cosh^2\!\rho\, ds^2_{\ads_d})~,
\eea
where $L$ is the AdS length scale and  $-\rho_{\star}<\rho<\infty$, with $\infty$ being the asymptotic boundary and $\rho_{\star}$ the location of  the brane.
In these coordinates, the extrinsic curvature takes the form $K_{ab}=-\frac{1}{2 L}\pderi{h_{ab}}{\rho}$ and \eqref{eq: junction} yields
\bea
\tanh \rho_{\star} = L T~.
\eea
Positive/negative tension implies positive/negative $\rho_{\star}$.

In the bottom-up model, the boundary conditions for geodesics reaching the brane have to be specified by hand, and a common choice is a reflective boundary condition, as e.g.\ in \cite{Reeves:2021sab,Omiya:2021olc,Kastikainen:2021ybu}.\footnote{In top-down BCFT duals the analogous conditions are dictated by the geometry and lead to a richer picture, which will be discussed in section \ref{sec: D1-D5}.}
Then geodesics launched from the conformal boundary of $\ads_{d+1}$ towards the brane can be reflected back to the conformal boundary. The conformal cross-ratio $\xi$ between the initial and final point on the asymptotic boundary can be determined from the geodesic equations without explicitly solving them \cite{Reeves:2021sab}.
Starting point is the metric \eqref{eq: ads metric slice}. With a coordinate transformation $\tanh (\frac{\rho}{2})=\tan (\frac{\theta}{2})$ and $\ads_d$ in global coordinates, $ds_{\ads_d}^2=\sec^2\!v(-dt^2+dv^2+\sin^2 v \,d\Omega_{d-2})$ with $|v|\leq\frac{\pi}{2}$ and conformal boundary at $|v|=\frac{\pi}{2}$, \eqref{eq: ads metric slice} becomes
\begin{equation}\label{eq: metric-ads-slice-gen}
	d s^2=\frac{L^2}{\cos^2\! \theta \cos^2\! v} \left[-d t^2+dv^2+\cos^2\! v\, d \theta^2+\sin^2\!v\, d\Omega_{d-2} \right]~,
\end{equation}
where $-\thes<\theta<\frac{\pi}{2}$. The brane is located at $\theta=-\thes=-\sin^{-1}(LT)$ and the asymptotic boundary at $\theta=\frac{\pi}{2}$.
One can focus on geodesics without angular momentum on $S^{d-2}$.\footnote{The return point for geodesics with angular momentum on $S^{d-2}$ is related by conformal symmetry, and leads to the same cross ratio as for geodesics without angular momentum.}
Null geodesics only depend on the conformal structure, so one can drop the conformal factor in their study.
The spatial slices $(v,\theta)$ are then simply a part of $2$-spheres.
Using the spherical law of cosines, the initial position $(v_0, \frac{\pi}{2})$, the return position $(v_1, -\frac{\pi}{2}-2\thes)$ and the elapsed time $\Delta t=d$ (the traveled geodesic distance on $S^2$) are related by \cite{Reeves:2021sab}
\bea
\cos d=\cos v_0 \cos v_1+\sin v_0 \sin v_1 \cos (2\thes+\pi)~.
\eea
and the cross ratio \eqref{eq: cross-ratio} is 
\bea \label{eq:cross-singular}
\xi_{\star}=-\cos^2 \thes=-\sech^2\rho_{\star}~.
\eea
In general $\xi_{\star}$ is neither of the values $-1,0$ allowed in the BCFT.
The exceptions are $\theta_{\star}\rightarrow -\frac{\pi}{2}$, corresponding to the near-critical limit, and $\theta_{\star}=0$, corresponding to a tensionless brane. In both cases the bulk-cone singularities coincide with expected BCFT singularities.

We note that the BCFT symmetries do not require the holographic dual to be locally AdS$_{d+1}$; they only impose the defect conformal symmetry realized by the $AdS_d$ slices. This allows for generic warp factors $f(\theta)^{2}$ instead of $L^2/\cos^{2}(\theta)$ in \eqref{eq: metric-ads-slice-gen}, which can be realized by adding background and dynamical fields to the action \eqref{eq: action-brane}. 
The arguments above extend to this case, since the causal structure does not depend on the conformal factor. The cross ratio thus remains the same for branes with reflective boundary condition.

\section{Penrose limit in braneworld models} \label{sec: Pen-brane}

A key ingredient in showing that the bulk-cone singularities at $\xi=\xi_\star$ (see \eqref{eq:cross-singular}) are resolved in string theory is the geometry in the vicinity of the null geodesics leading to the bulk-cone singularities, which is described by the associated Penrose limit.

We focus in this section on the simplest bottom-up models where an $\ads_{d+1}$ spacetime with metric \eqref{eq: ads metric slice} is truncated by an ETW brane with constant tension. A key simplification in this model results from the fact that Penrose limits of AdS spacetimes are flat \cite{Blau:2024,Siopsis:2002vw}. The geometry explored by geodesics scattering off the ETW brane is AdS almost everywhere, except for the actual reflection at the brane. Therefore, we only expect a non-trivial pp-wave geometry near the reflection point on the ETW brane. 

To determine the form of the pp-wave geometry more explicitly, it is useful to double the geometry (\ref{eq: metric-ads-slice-gen}) by extending it past the ETW brane, so that the metric takes the form
\begin{subequations}\label{eq: metric-double}
	\begin{equation}
		d s^2=\frac{L^2}{g^2 ( \theta) \cos^2\! v} \left[-dt^2+dv^2+\cos^2\!v\, d \theta^2+\sin^2\!v\, d\Omega_{d-2}\right],
	\end{equation}
	with doubled azimuth angle $-\frac{\pi}{2}-2\theta_\star<\theta<\frac{\pi}{2}$ and
	\begin{equation} 
		g(\theta)= 
		\begin{cases}
			\cos \theta & -\thes<\theta<\frac{\pi}{2}  \\ 
			\cos (\theta+2\thes) & -\frac{\pi}{2}-2\thes<\theta<-\thes\end{cases}~.
	\end{equation}
\end{subequations}
Identifying $\theta\sim -2\theta_\star - \theta$ recovers the original geometry. In the doubled geometry, a geodesic reflected off the ETW brane back into the bulk lifts to a geodesic continuing smoothly into the region $\theta<-\theta_\star$, while experiencing a kink in the warp factor $g(\theta)$.

The next ingredient are the null geodesics themselves, where we focus, as in section~\ref{sec: bulk-cone BCFT}, on geodesics without angular momentum on $S^{d-2}$. They are parametrized by $(t(\lambda),v(\lambda),\theta(\lambda))$ with affine parameter $\lambda$ and the geodesic equations are
\begin{align} \label{eq: geo-ads3}
	-t'^2+v'^2+\cos^2 v \,\theta'^2 &=0~, &&\text{null condition}
	\nonumber\\
	\cos^2 v \,\dot{\theta}&=\ell~, &&\text{conserved $\theta$ momentum}
	\nonumber\\
	\frac{t'}{g^2(\theta)\cos^2 v }&=c_1~, &&\text{conserved $t$ energy}
\end{align}
where $\cdot$ and $'$ denote derivatives with respect to $t$ and the affine parameter, respectively.
The momentum satisfies $\ell^2<1$ and we set the constant $c_1=1$ by rescaling the affine parameter. The solutions are, with integration constants $\theta_0$, $t_0$,
\begin{align}\label{eq: geodesic}
	\sin^2v&=(1-\ell^2)\sin^2(t-t_0)~,
	&
	\theta&=\theta_0+ \tan^{-1}\left(\ell\tan(t-t_0)\right)~.
\end{align}

We now turn to the Penrose limit. We refer to \cite{Blau:2024} for a broad overview, and to appendix~\ref{appendix: Penrose} for a review of some background relevant here.
In order to take the Penrose limit we perform coordinate transformations in two steps. In the first step, we define
\begin{align}\label{eq: adapted-ads3}
	\frac{d\hat V}{L}  & = -dt + \dot{v}\,dv + \ell \,d\theta~,  
	\nonumber\\
	\frac{dU}{L} & = \frac{d\theta}{\ell g^2(\theta) }~,  
	\nonumber\\
	\frac{d\hat W}{L} & = -\frac{\ell^2}{\dot{v}\, \cos^2 v}\,dv +\ell\,d\theta~.
\end{align}
By \eqref{eq: geo-ads3}, $\dot{v}^2=1-\ell^2/\cos^2\!v$, so the coefficients of $(dt,dv,d\theta)$ are all functions of the respective coordinates and the differential forms are exact. In particular, we can solve
\begin{align}\label{eq:U-theta}
	U&=\begin{cases}
		- \frac{L}{\ell}\tan (\theta) & -\thes<\theta<\frac{\pi}{2}\\
		-  \frac{L}{\ell}\tan (\theta+2\thes)+2U_{\star} & -\frac{\pi}{2}-2\thes<\theta<-\thes
	\end{cases}~,
\end{align}
where $U_{\star}=\frac{L}{\ell}\tan (\thes)$.
In these coordinates $g_{UU}=g_{U\hat W}=0$ and $U$ can be identified with the affine parameter for the null geodesics.
The Penrose limit amounts to the rescaling $(U,\hat V,\hat W,L^2d\Omega_{d-2}^2)\to(U,a^2 \hat V,a \hat W,a^2ds^2_{\mrr^{d-2}})$\footnote{The sphere on which the geodesic is stationary,  $d\Omega_{d-2}^2=d\vartheta^2+\sin^2\!\vartheta\, d\Omega_{d-3}^2$, becomes in the Penrose limit, where $\vartheta\to a \vartheta$ with $a\to 0$, flat space in polar coordinates, $d\Omega_{d-2}^2\to a^2(d\vartheta^2+\vartheta^2d\Omega_{d-3}^2)=a^2ds^2_{\RR^{d-2}}$.} and the limit $a\to 0$,  leading to the pp-wave geometry 
\begin{subequations}\label{eq: pp-wave-rosen}
\begin{equation}
	ds^2  = 2dU d\hat V+e^2 (U)d\hat W^2+\er^2(U) ds^2_{\mrr^{d-2}}~,
\end{equation}
where
\begin{equation}
	e_W^2 (U)=\frac{1-\ell^2\sec^2\! v}{\ell^{2}g^{2}(\theta)}~,
	\qquad\qquad
	\er^2 (U)=\frac{\tan^2\!v}{g^{2}(\theta)}~.
\end{equation}
\end{subequations}
In these expressions $\theta(U)$ and $v(U)$ take the on-shell forms for the null geodesics, which can be obtained from $U(\theta)$ in (\ref{eq:U-theta}) and $(\theta(t),v(t))$ in \eqref{eq: geodesic}. This leads to
\begin{align}\label{eq:e-sq}
	e_W^2(U)&=\frac{1-\ell^2}{\ell^2L^4}\begin{cases}\left(
		(L^2+\ell^2 U U_0)\cos\theta_0 \right)^2
		& -\thes<\theta\\
		\left((L^2+\ell^2 U\tilde U_0)\cos(\theta_0+2\theta_\star)-2(L^2+\ell^2 U_\star  U_0)\cos\theta_0 \right)^2
		&-\thes>\theta
	\end{cases}\,,
	\nonumber\\
	e_{\mrr}^2(U)&=\frac{1-\ell^2}{L^2}\begin{cases}
		\left( (U - U_0)\cos\theta_0\right)^2~ 
		& -\thes<\theta\\
		\left(
		(\tilde U_0-U) \cos(\theta_0 + 2\theta_\star)
		- 2(U_{\star}-U_0)\cos\theta_0
		\right)^2
		&-\thes>\theta
	\end{cases}\,,
\end{align}
where the constants $U_0$ and $\tilde U_0$ are defined as
\begin{align}
	U_0&=-\frac{L}{\ell}\tan\theta_0~,
	&
	\tilde U_0&=+\frac{L}{\ell}\tan(\theta_0+2\thes)~.
\end{align}
Similar to the warp factor $g^2(\theta)$, $e_{W/\RR}^2(U)$ are piecewise smooth with a kink at the brane.
In a second step we bring the metric to the form \eqref{eq:pp-wave-0} by a further coordinate transformation
\begin{align}
	\hat V&=V+\frac{1}{2}\frac{e_W'(U)}{e_W(U)}W^2+\frac{1}{2}\frac{e_\mrr'(U)}{e_\mrr(U)}\hat{x}^2~, & \hat W&=\frac{W}{e_W(U)}~,&\hat{x}=\frac{x}{e_\mrr (U)}~,
\end{align}
where $\hat{x}$ denotes the  Cartesian coordinate on $\mrr_{d-2}$. This leads to 
\begin{subequations}\label{eq: pp-wave-3d}
\begin{equation}
	ds^2=2dUdV+(A_{WW}(U) W^2+A_{\mrr}(U)x^2) dU^2+ds^2_{\mrr^{d-1}}~,
\end{equation}
where
\begin{equation}
	A_{WW}(U)=\frac{e_W''(U)}{e_W(U)}~,
	\qquad\qquad
	A_{\mrr}(U)=\frac{e_\mrr''(U)}{e_\mrr(U)}~.
\end{equation}
\end{subequations}
This form of the metric will be used for the string theory discussion.
The non-trivial information is in the pp-wave profiles $A_{WW}$ and $A_{\mrr}$.

The remaining task is to determine $A_{WW}$ and $A_{\mrr}$. To avoid square roots one can use $e''/e=\frac{1}{2}(e^2)''/e^2-\frac{1}{4}((e^2)')^2/e^{4}$. This is equivalent to choosing the branches such that
\begin{align}\label{eq:eU}
	e_W(U)&=\frac{\sqrt{1-\ell^2}}{\ell L^2}\begin{cases}
		(L^2 + \ell^2U U_0 )\cos\theta_0
		& -\thes<\theta
		\\
		-(L^2 + \ell^2 U \tilde{U}_0 )\cos(\theta_0 + 2\theta_{\star})
+ 2 (L^2 + \ell^2 U_{\star} U_0 )\cos\theta_0
		& -\thes>\theta
	\end{cases}~,
	\nonumber\\
e_\mrr(U)&=\frac{\sqrt{1-\ell^2}}{ L}\begin{cases}
		(U - U_0)\cos(\theta_0)
		& -\thes<\theta\\
	-(\tilde{U}_0 - U)\cos(\theta_0 + 2\theta_{\star})
+ 2 (U_{\star} - U_0)\cos(\theta_0)
		& -\thes>\theta
	\end{cases}~.
\end{align}
Both are piecewise linear with kinks at the brane, where $e_{W}(U_{\star})=\frac{\sqrt{1 - \ell^2}}{\ell} \cos(\theta_0 + \theta_{\star}) \sec(\theta_{\star})$ and $e_\mrr(U_{\star})=\frac{\sqrt{1 - \ell^2}}{\ell} \sin(\theta_0 + \theta_{\star})\sec(\theta_{\star})$.
As a result, $A_{WW}$ and $A_\RR$ vanish where $e_{W/\RR}(U)$ are smooth, reflecting that the Penrose limit of undeformed AdS is flat.
The derivatives of $e_{W/\RR}(U)$ have to be understood as distributions, with $e_{W/\RR}'(U)$ piecewise constant and $e_{W/\RR}''(U)$ $\delta$-functions. They can be determined from (\ref{eq:eU}), leading to
\begin{align}\label{eq: A-brane}
	A_{WW}(U)&=A_{\mrr}(U)=A\,\delta(U-U_\star)~, &
	A&=-\frac{\ell}{L} \sin(2\thes)~.
\end{align}
We note that $A>0$ for negative tension branes and $A<0$ for positive tension branes.
For the tensionless brane $\thes=0$ and the near-critical limit $\thes\rightarrow \frac{\pi}{2}$, $A$ vanishes.

\section{Resolution of bulk-cone singularities}\label{sec: resolve}

We now turn to string theory $\ap$-corrections to the two-point function with kinematics exhibiting the bulk-cone singularities. In this context we should address that braneworld models are bottom-up constructions: the ETW-branes are added by hand, with no clear string theory origin -- the models do not arise as consistent truncations from string theory and have no clear uplift (except for the tensionless case). They should be seen as useful toy models rather than full-fledged theories of quantum gravity. We will nevertheless proceed with string theory. Operationally, we focus on the spacetime dimensions described by the braneworld model, and in particular on the Penrose limit describing the geometry near the null-geodesics responsible for the bulk-cone singularities. The latter yields pp-wave geometries of the form (\ref{eq: metric-shock-pp-pre}), as shown in section~\ref{sec: Pen-brane}. This means we only need the pp-wave to have an embedding into string theory, not the entire braneworld model. This is a milder assumption. We will show in section \ref{sec: D1-D5} that such shockwave pp-waves indeed arise in certain regimes of Penrose limits in top-down BCFT duals.

We will show that this approach of incorporating stringy $\ap$-corrections into the bottom-up models, in particular to the two-point function with kinematics exhibiting the bulk-cone singularity, leads to an upper bound and thus resolves the bulk-cone singularities.

\subsection{Quantization of strings and particle production}\label{sec: string theory}

As a first step we quantize string theory on the pp-wave background \eqref{eq: pp-wave-3d} with pp-wave profile \eqref{eq: A-brane}. 
The pp-wave profile leads to a time-dependent mass for the worldsheet fields, and we compute the particle production resulting from this shockwave form. 
We focus on the 3-dimensional part of the geometry parametrized by $(U,V,W)$, which was relevant for the geodesics discussed in section \ref{sec: Pen-brane}, so that the relevant geometry takes the form
\begin{align}\label{eq:ds2-eff-string}
	ds^2&=2dUdV+A\delta(U-U_\star) W^2 dU^2+dW^2~, & A&=-\frac{\ell}{L} \sin(2\thes)~.
\end{align}
Strings on similar (but different) shockwave backgrounds have been studied in \cite{deVega:1988ts,Giddings:2007bw}.

Starting point is the bosonic part of string worldsheet action in general curved spacetime, as in \cite{Horowitz:1990sr},
\begin{equation} \label{eq: action-string-gen}
	\begin{gathered}
		S=-\frac{1}{4 \pi \alpha^{\prime}} \int\left(h^{a b} g_{\mu v} \partial_a X^\mu \partial_b X^v+B_{\mu v} \partial_a X^\mu \partial_b X^v \epsilon^{a b}\right. 
		\left.-\frac{1}{2} \alpha^{\prime} R^{(2)} \Phi\right) \sqrt{h} d^2 \sigma~,
	\end{gathered}
\end{equation}	
where $B_{\mn}$ is the NS-NS two-form and $\Phi$ the dilaton.
Following the above discussion, we drop $B_{\mu\nu}$ and the dilaton, and work with toy models without those fields.
The action on a general pp-wave background of the form \eqref{eq:pp-wave-0} in the light-cone gauge $U=\ap p_v \tau$ is
\begin{align}\label{eq: action-gf}
	S=-\frac{1}{4 \pi \alpha^{\prime}} \int d^2 \sigma
	\sum_{\alpha,\beta}\left(-(\p_{\tau} X^{\alpha})^2+(\p_{\sigma} X^{\alpha})^2-(\ap p_v)^2 A_{\alpha\beta}(\ap p_v\tau) X^{\alpha} X^{\beta}\right)~,
\end{align}
where $X^{\alpha}$ denote the transverse directions. 
The equations of motion for the transverse fields are linear, with time-dependent mass determined by the pp-wave profile $A_{\alpha\beta}$,
\begin{equation}\label{eq: string eom pp gen}
	\left(\partial_\tau^2-\partial_\sigma^2\right) X^{\alpha}(\tau,\sigma)=\sum_{\beta} (\ap p_v)^2 A_{\alpha\beta}\left(\ap p_v \tau\right) X^{\beta}(\tau,\sigma)~.
\end{equation}
The constraint resulting from vanishing of the world-sheet stress tensor can be solved for $V(\tau,\sigma)$ in terms of the transverse fields $X^{\alpha}(\tau,\sigma)$.

For the effective 3-dimensional shockwave pp-wave geometry in (\ref{eq:ds2-eff-string}) we use the gauge $U=U_\star+\ap p_v \tau$, which shifts $\tau$ conveniently. There is only one transverse direction $W$, for which the equation of motion in \eqref{eq: string eom pp gen} becomes
\be\label{eq:W-eom}
\lbra{\p_{\tau}^2-\p_{\sigma}^2}W(\tau,\sigma)-\ap p_v A \delta(\tau) W(\tau,\sigma)=0~.
\ee
The strings are free away from $\tau = 0$ with a $\delta$-function source at $\tau = 0$.
As a result, $W(\tau,\sigma)$ is continuous at $\tau=0$ while its time derivative jumps,
\bea
\wo(0,\sigma)&=\wi(0,\sigma)~,\\
\lmid{\p_{\tau}\lbra{\wo(\tau,\sigma)-\wi(\tau,\sigma)}}\Big\vert_{\tau=0}&=\ap p_v A W(0,\sigma)~.
\eea
The solution is given by
\bea \label{eq:sol-sandwich}
\wo(\tau,\sigma)=\wi(\tau,\sigma)
+\frac{\ap p_v A}{2}\int_{\sigma-\tau}^
{\sigma+\tau}d\sigma' W(0,\sigma')~.
\eea
We now compute the Bogoliubov coefficients and particle production.
First we expand the free outgoing string solutions ($\tau>0$) and ingoing solutions ($\tau<0$) in terms of creation and annihilation operators,
\begin{align}
	W_{\mathrm{out}}(\tau,\sigma) &= x_0^{\mathrm{out}} + 2 \alpha^{\prime} p^{\mathrm{out}} \tau + i \sqrt{\frac{\alpha^{\prime}}{2}} \sum_{n \neq 0} \frac{1}{n} \left[ \alpha_n^{\mathrm{out}} e^{-i n(\tau+\sigma)} + \tilde{\alpha}_n^{\mathrm{out}} e^{-i n(\tau-\sigma)} \right],
	\nonumber\\
	W_{\mathrm{in}}(\tau,\sigma) &= x_0^{\mathrm{in}} + 2 \alpha^{\prime} p^{\mathrm{in}} \tau + i \sqrt{\frac{\alpha^{\prime}}{2}} \sum_{n \neq 0} \frac{1}{n} \left[ \alpha_n^{\mathrm{in}} e^{-i n(\tau+\sigma)} + \tilde{\alpha}_n^{\mathrm{in}} e^{-i n(\tau-\sigma)} \right].
\end{align}
Using \eqref{eq:sol-sandwich}, we find the Bogoliubov transform
\begin{equation}
	\begin{aligned}
		&\alpha_n^{\mathrm{out}} = \left( 1 + \frac{i \alpha' A p_v}{2n} \right) \alpha_n^{\mathrm{in}} - \frac{i \alpha' A p_v}{2n} \tilde{\alpha}_{-n}^{\mathrm{in}}~, \\
		&\tilde{\alpha}_n^{\mathrm{out}} = \left( 1 + \frac{i \alpha' A p_v}{2n} \right) \tilde{\alpha}_n^{\mathrm{in}} - \frac{i \alpha' A p_v}{2n} \alpha_{-n}^{\mathrm{in}}~.
	\end{aligned}
\end{equation}
The occupation numbers for the outgoing modes in the ingoing vacuum are
\bea
\bra{0_{\mathrm{in}},p_v}N_n^{\mathrm{out}} \ket{0_{\mathrm{in}},p_v}=\lbra{\frac{ \alpha' A p_v}{2n}}^2~.
\eea
We can also compute the overlap between the early and late vacua \cite{DeWitt:1975ys}
\bea\label{eq: overlap-vacuum}
\bra{\mathrm{out}, p_v} \ket{\mathrm{in}, p_v}
&=\prod_{n=1}^{\infty}\left(1+\left\langle N_n^{\text{out}}\right\rangle\right)^{-1 / 2}\\
&=\prod_{n=1}^{\infty}\left(1+\lbra{\frac{ \alpha' A p_v}{2n}}^2\right)^{-1 / 2}
=\sqrt{\frac{\pi\,\alpha' A p_v}{2\,\sinh \bigl(\tfrac{\pi\,\alpha' A p_v}{2}\bigr)}}\,. 
\eea 
where the last equation follows from $\sin(\pi z) \;=\; \pi z \prod_{n=1}^\infty (1 - z^2/n^2)$ for complex $z$.
In particular, we note that the overlap is exponentially suppressed at large $p_v$.

\subsection{Resolution of bulk-cone singularities}
Now we discuss the two-point functions.
The stringy version of the bulk-to-bulk two-point function is a curved-space generalization of flat-space closed string amplitudes with Dirichlet boundary conditions \cite{Cohen:1985sm,Green:1987mn}.
Following \cite{Cohen:1985sm}, we start with the amplitude $\mathcal{D}(\ell_i,\ell_f)$ for a closed string localized at a loop $\ell_i$ at $\tau=\tau_i$ to be detected at a loop $\ell_f$ at $\tau=\tau_f$,
\bea
\mathcal{D}(\ell_i,\ell_f)=\int d\Sigma \int \frac{Dh}{V_{\text{diff}\times\text{Weyl}}}\int_{X(\tau_i, \sigma)=\ell_i}^{X(\tau_f, \sigma)=\ell_f} DX e^{i S_p}~,
\eea
where $h$ denotes the world-sheet metric, $X(\tau,\sigma)$ denotes the world-sheet scalars, and $S_p$ is the Polyakov action as in \eqref{eq: action-string-gen}.
$d\Sigma$ represents the integral over reparameterizations of the loops in the boundary conditions $X(\tau_{i,f},\sigma)=\ell_{i,f}$.
The functional measures $DX$, $Dh$ and the division by the gauge groups are standard in the Polyakov path integral. 
Fixing the gauge symmetry by imposing light-cone gauge, as in section~\ref{sec: string theory}, the amplitude becomes
\bea
\mathcal{D}(\ell_i,\ell_f)=\int d\Sigma\int_{X(\tau_i, \sigma)=\ell_i}^{X(\tau_f, \sigma)=\ell_f} DX^{\alpha} e^{i S_{gf}}~,
\eea
where $S_{gf}$ is the gauge-fixed action in \eqref{eq: action-gf}, and we are left with the transverse modes $X^{\alpha}$.
We focus on the special case where $\ell_{i,f}$ shrink to points $x_{i,f}$ to connect to two-point functions in quantum field theory.
In this case, $X(\tau_{i,f},\sigma)=x_{i,f}$ are constant maps and the $\Sigma$ integral drops out.
This leads to the stringy version of the bulk-to-bulk two-point function between $x_i=(u_i,v_i,x_i^{\alpha})$ and $x_f=(u_f,v_f,x_f^{\alpha})$,\footnote{To motivate this amplitude as stringy version of quantum field theory two-point functions, it was shown in \cite{Cohen:1985sm} that in flat space this amplitude is a product of two-point functions of the particles in the infinite stringy tower. Boundary two-point functions are obtained by taking the boundary limit of the bulk two-point functions, and subtleties in this procedure were discussed in \cite{Dodelson:2020lal}.} which was also interpreted as annulus amplitude with two $D({-1})$ branes at the two bulk points in \cite{ Dodelson:2020lal}.
We Fourier-transform the $v$ coordinate to momentum space, with momentum denoted $p_v$, and use the notation
\begin{align}\label{eq:prop}
	G\left(u_f, u_i, v_f-v_i, x_f^{\alpha}, x_i^{\alpha}\right)=
	\mathcal{D}(x_i,x_f)&=\int_{-\infty}^{\infty} d p_v e^{i p_v\left(v_f-v_i\right)}\left\langle p_v, x_f^{\alpha}, u_f \mid p_v, x_i^{\alpha}, u_i\right\rangle,
\end{align}
where
\begin{align}
	\left\langle p_v, x_f^{\alpha}, u_f \mid p_v, x_i^{\alpha}, u_i\right\rangle & =\int_{X^{\alpha}\left(\tau_i, \sigma\right)=x_i^{\alpha}}^{X^{\alpha}\left(\tau_f, \sigma\right)=x_f^{\alpha}} D X^{\alpha} e^{i S_{gf}}~.
\end{align}
We further split $X^{\alpha}(\tau,\sigma)$ into the center-of-mass string motion and the transverse string fluctuations. So the path integral splits into a zero-mode propagator $G_0$ and a determinant resulting from the transverse fluctuations with quadratic action,
\begin{equation}\label{eq: stringy-two-point}
	\begin{aligned}
		\int_{X^{\alpha}\left(\tau_i, \sigma\right)=x_i^{\alpha}}^{X^{\alpha}\left(\tau_f, \sigma\right)=x_f^{\alpha}} D X^{\alpha} e^{i S\left[X^{\alpha}\right]} 
		=\frac{G_0\left(p_v, u_f, u_i, x_f^{\alpha}, x_i^{\alpha}\right)}{\prod_{\alpha} \prod_{n=1}^{\infty} \operatorname{det}\left(-\partial_\tau^2-n^2+p_v^2 A_{\alpha\alpha}\left(p_v \tau\right)\right)}~.
	\end{aligned}
\end{equation}
The determinants are evaluated with the boundary conditions specified in the path integral, and we project onto the initial and final vacuum states, taking $\tau_{i/f} \to \tau_{i/f}(1 - i\epsilon)$ and then $\tau_f \to \infty$ and $\tau_i \to -\infty$.
The momentum-space propagator becomes
\begin{equation}\label{eq:G0-split}
	\begin{aligned}
		\left\langle p_v, x_f^{\alpha}, u_f \mid p_v, x_i^{\alpha}, u_i\right\rangle 
		& =G_0\left(p_v, u_f, u_i, x_f^{\alpha}, x_i^{\alpha}\right) \bra{\mathrm{out}, p_v} \ket{\mathrm{in}, p_v}\,,
	\end{aligned}
\end{equation}
where the overlap between the vacua before and after the reflection at the ETW brane was computed in (\ref{eq: overlap-vacuum}).
The zero-mode propagator near the light cone in $D$ dimensions can be expressed as \cite{DeWitt:1975ys} (see also \cite{Dodelson:2020lal})
\begin{equation}
	G_0\left(p_v, u_f, u_i, x_f^{\alpha}=x_i^{\alpha}=0\right)=\frac{\Delta^{1/2}(u_f,u_i)}{\left((u_f-u_i)(v_f-v_i)+i\epsilon\right)^{D/2-1}}~,
\end{equation}
where $\Delta$ is the Van-Vleck determinant, coming from the Gaussian integral around the classical trajectory (the null geodesic leading to the bulk-cone singularity); it is independent of $p_v$ and finite. Fourier-transforming leads to 
\begin{equation}\label{eq:G0-p}
	G_0\left(p_v, u_f, u_i, x_f^{\alpha}=x_i^{\alpha}=0\right)=
	C(p_v)^{D/2-2} \,\Theta\left(p_v\right)\,,
\end{equation}
with a $p_v$-independent $C$  given by
\begin{equation}
	C=\frac{i^{-\frac{D}{2}+1}}{\Gamma(D/2-1)}\frac{\Delta^{1 / 2}\left(u_f, u_i\right)}{\left(u_f-u_i\right)^{D/2-1}}\,.
\end{equation}

Finiteness of the propagator on the light cone can now be shown starting from (\ref{eq:prop}) and using the triangle inequality, 
\begin{align}\label{eq: uppder-bound}
	\left|G\left(u_f, u_i, v_f-v_i=x_i^{\alpha}=x_f^{\alpha}=0\right)\right| & \leq \int_{-\infty}^{\infty} d p_v\left|\left\langle p_v, x_f^{\alpha}, u_f \mid p_v, x_i^{\alpha}, u_i\right\rangle\right| 
	\nonumber\\
	&=\int_{-\infty}^{\infty} d p_v \abs{G_0\left(p_v, u_f, u_i, x_f^{\alpha}=x_i^{\alpha}=0\right)\bra{\mathrm{out}, p_v} \ket{\mathrm{in}, p_v} }
	\nonumber\\
	&=C \int_0^{\infty}dp_v\,(p_v)^{D/2-2}\sqrt{\frac{\pi\,\alpha' A\,p_v}{2\,\sinh \left(\frac{\pi\,\alpha' A }{2}p_v\right)}}~.
\end{align}
The second line follows from (\ref{eq:G0-split}), the third from (\ref{eq:G0-p}) and (\ref{eq: overlap-vacuum}).
For $D>3$ the integrand is finite; for $D=3$ it is singular but integrable at $p_v=0$.
The integral in the third line is finite for $D\geq 3$ due to the exponential suppression at large momentum resulting from the stringy worldsheet particle production. This shows that the stringy propagator is finite.
The right hand side in the last line of (\ref{eq: uppder-bound}) can be evaluated explicitly, to give
\begin{align}\label{eq:bound-expl}
	\left|G\left(u_f, u_i, v_f-v_i=x_i^{\alpha}=x_f^{\alpha}=0\right)\right| & \leq \frac{\tilde C}{(\ap \abs{A})^{D/2-1}}~,
\end{align}
where a dimension-dependent constant has been absorbed into $C$, resulting in $\tilde C$.
We recall that the constant $A$, given in (\ref{eq:ds2-eff-string}), depends on the brane tension through $\thes$.

The constant $A$ on the right hand side in (\ref{eq:bound-expl}) vanishes for $\thes=0$ and $\thes=\frac{\pi}{2}$, leading to a diverging bound. In these cases even the stringy two-point functions are not bounded by our considerations. The case $\thes=0$ corresponds to a tensionless brane. In this case the bottom-up geometry can be understood as an orbifold of $\ads_{d+1}$, and the bottom-up models have an uplift, e.g.\ to orbifolds of $\ads_5\times {\rm S}^5$ in Type IIB. As remarked at the end of section \ref{sec: review-sin}, the bulk-cone singularity in this case coincides with expected BCFT singularities and does not need resolution. In the near-critical limit $\thes\rightarrow\frac{\pi}{2}$ the bulk-cone singularities likewise coincide with expected BCFT singularities and do not need resolution.

To close the discussion on braneworld models, we recall that, in general, braneworld duals for BCFTs should incorporate non-trivial one-point functions and allow deformed bulk geometries which are only asymptotically AdS$_{d+1}$. This amounts to allowing more general warp factors $f(\theta)$ instead of $L/\cos \theta$ e.g.\ in \eqref{eq: metric-ads-slice-gen}, so that the geometry becomes
\begin{equation}\label{eq:ds2-braneworld-gen}
	ds^2_{d+1}=f(\theta)^2\left(ds^2_{\ads_d}+d\theta^2\right)~,
\end{equation}
with an ETW brane terminating the geometry at some $\thes$.
Geodesics leading to bulk-cone singularities can still be found, as discussed at the end of section~\ref{sec: bulk-cone BCFT}. But they now lead to pp-wave profiles which are non-trivial also in the bulk, while still featuring a shock due to the ETW-brane. The above arguments can be extended to this case. To show this in a concrete example, we will turn to a top-down model with a concrete geometry: the discussion in section~\ref{sec:large-p-top-down} will show that the above arguments still apply to the large-$p_v$ regime in the more general braneworlds, which is sufficient to bound the two-point function.

\section{Top-down holography: D1/D5 BCFTs}\label{sec: D1-D5}

In this section we study top-down holographic duals for 2d BCFTs based on the D1/D5 system. This will serve to illustrate general features, which are also shared e.g.\ by holographic duals for 4d $\mathcal N=4$ SYM BCFTs.
We will first introduce the setup, then contrast the features of geodesics with bottom-up models, and discuss bulk-cone singularities.

\subsection{$\ads_2$ cap solutions}\label{eq:cap-sol}

\begin{figure}
	\centering
	\subfigure[][]{\label{fig:D1D5-junction}
		\begin{tikzpicture}
			\draw[thick] (-1,1) -- (0,0.02) -- (1.6,0.02);
			\draw[thick] (-1,-1) -- (0,-0.02) -- (1.6,-0.02);
			\node [anchor=west] at (-1,1) {\, D1};
			\node [anchor=west] at (-1,-1) {\, D5};
			\node [anchor=south] at (1,0) {D1/D5};
		\end{tikzpicture}
	}
	\hskip 20mm
	\subfigure[][]{\label{fig:AdS2xS2-sol-w}
		\begin{tikzpicture}[scale=1.1]
			\draw[color=white,pattern=north west lines, pattern color=gray] (-1.95,0) rectangle (1.95,1.45);
			\draw [thick,->] (-2,0) -- (2,0);
			\draw [thick,->] (0,0) -- (0,1.5);
			\draw[thick] (-1.2,-0.05) -- (-1.2,0.05);
			\draw[thick] (1.2,-0.05) -- (1.2,0.05);
			\node [anchor=north] at (1.2,0) {\footnotesize D1};
			\node [anchor=north] at (-1.2,0) {\footnotesize D5};
			\node [anchor=north] at (0,0) {\footnotesize D1/D5};
			\node at (1.5,1.2) {$\Sigma$, $w$};
		\end{tikzpicture}
	}
	\caption{Left: Brane construction with D1 and D5 joining to form a D1/D5 BCFT, proposed in \cite{Chiodaroli:2011fn}. Right: Supergravity solution on the upper half plane with coordinate $w$. The external branes are represented as a D1/D5 source at $w=0$, where the geometry becomes $\rm AdS_3\times S^3$, and separate D1 and D5 sources at $w=+1$ and $w=-1$.\label{fig:D1D5-brane}}
\end{figure}
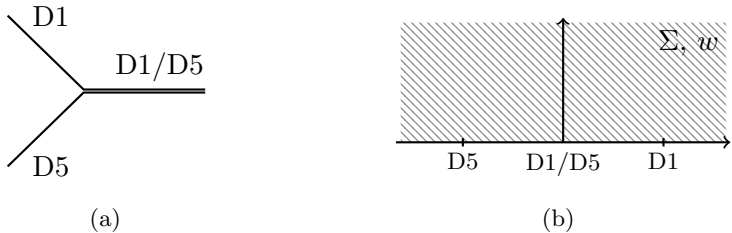

We focus on BCFTs based on D1/D5 brane setups, described holographically by solutions constructed in \cite{Chiodaroli:2011nr,Chiodaroli:2011fn,Chiodaroli:2012vc} in 6d type 4b gauged supergravity coupled to a number $m$ of tensor multiplets. This 6d theory arises from Type IIB compactified on $T^4$ or $K3$, with $m=5$ for $T^4$ and $m=21$ for $K3$ \cite{Chiodaroli:2012vc}. 
The $m=5$ theory is a consistent truncation of Type IIB on $T^4$ (this follows from a symmetry argument: the zero modes on $T^4$, which are invariant under the torus isometries, can not source modes which transform non-trivially). The 6d geometry is a warped product of $\rm AdS_2$ and $\rm S^2$ over a Riemann surface $\Sigma$, with line element\footnote{Note that $ds^2_\Sigma=4\rho^2|dw|^2$ in (3.1), (3.22) of \cite{Chiodaroli:2011nr}; the factor 4 was omitted below (2.2) of \cite{Chiodaroli:2011fn} and seems to have been lost in \cite{Reeves:2021sab}, but it is needed e.g.\ to get $\rm AdS_3\times S^3$ at $w=0$. \label{foot:rho2-conv-1}}
\begin{align}\label{eq:metric}
	ds^2&=f_1^2 ds^2_{\rm AdS_2}+f_2^2 ds^2_{\rm S^2}+4\rho^2 |dw|^2~.
\end{align}
It is supported by fluxes and other non-trivial 6d supergravity fields, whose explicit expressions we will not use.
We focus on a particular solution, called  $\rm AdS_2$ cap in \cite{Chiodaroli:2011fn}, where $\Sigma$ is the upper half plane with complex coordinate $w$ and the metric functions are
\begin{align}\label{eq:f1f2rho-w}
	f_1^4&=\frac{2\kappa^2 \cY}{\mu_0^2|w|^4}~,
	&
	f_2^4&=\frac{2\kappa^2\mu_0^2\Im(w)^4}{|w|^4\cY}~,
	&
	\rho^4&=\frac{\mu_0^2\cY}{8\kappa^2|w|^4|1-w^2|^4}~,
\end{align}
where $\kappa$ and $\mu_0$ are real parameters and
\begin{align}\label{eq:cY}
	\cY&=\mu_0^2\Im(w)^2+\kappa^2|1-w^2|^2~.
\end{align}
The solutions have a smooth, closed internal space aside from singularities at $w\in\{0,\pm 1\}$: the $\rm S^2$ collapses on the boundary of $\Sigma$ (the real line) with $f_2^2\to 0$, closing off the internal space smoothly, and $\rho$ falls off for large $|w|$ so $\Sigma$ can be compactified.

The $\ads_2$ cap solutions exist in both the $m=5$ and $m=21$ 6d supergravities, connecting them to $T^4$ and $K3$ compactifications of Type IIB.
They were interpreted as describing junctions of D1-branes with D5-branes wrapping $T^4$ or K3 in the 10d geometry, so as to realize a 2d CFT on half space. 
The solutions are illustrated in fig.~\ref{fig:D1D5-brane}.

At $w=0$ there is a combined D1/D5 source and an asymptotic $\rm AdS_3\times S^3$ region emerges.
The metric becomes, with $w=e^{-x+i\phi}$ and $x$ large \cite[(2.9),(4.11)]{Chiodaroli:2011fn}, 
\begin{align}\label{eq:metric-AdS3S3}
	ds^2&=\sqrt{2\mu_0^2}\left(d\phi^2+\sin^2\!\phi\, ds^2_{S^2}+dx^2+e^{2x}\frac{\kappa^2}{\mu_0^2}ds^2_{\rm AdS_2}\right)~.
\end{align}
This is $\ads_3\times S^3$ with radius set by $\mu_0^2$, which corresponds to the central charge $\sim N_{\rm D1}N_{\rm D5}$ of the D1/D5 CFT.
The conformal boundary of this $\ads_3\times S^3$ region is $\ads_2$, which is conformal to a half space. So the dual field theory is a 2d BCFT.
One of the ``$\rm AdS_2$ caps" at $w=\pm 1$ represents a D1-brane source and the other a D5-brane source.

\subsection{Resolved end-of-the-world branes are sticky}

The top-down solutions can be compared to the braneworld models in fig.~\ref{fig:AdS-brane} for $d=2$. We recall that one should allow for a general asymptotically-AdS$_3$ geometry as bulk in fig.~\ref{fig:AdS-brane} (see footnote \ref{foot:ads}). The bottom-up model can then be viewed as simplified description of the D1/D5 dual with the loose identifications\footnote{For holographic duals of BCFTs based on 4d $\mathcal N=4$ SYM a sharper identification was derived in \cite{Coccia:2021lpp}.}
between fig.~\ref{fig:AdS-brane} and fig.~\ref{fig:AdS2xS2-sol-w}:
\begin{align}
	&\text{asymptotically-$\ads_3$ bulk} & &\leftrightarrow && \text{asymptotic $\ads_3\times S^3$ at $w=0$ }
	\nonumber\\
	&\text{ETW brane} &&\leftrightarrow && 
	\text{6d geometry with $w=0$ removed}
	\nonumber
\end{align}
The distance from interior points of $\Sigma$ to $w=0$ loosely corresponds to the coordinate $\rho$ in the braneworld model. As explained below (\ref{eq:cY}), the top-down geometry closes off smoothly. Instead of being terminated abruptly by an ETW-brane, the $\ads_3\times S^3$ in the top-down setup ends in a smooth geometry supported by fluxes.

In the braneworld models, the geodesics leading to bulk-cone singularities are launched from the BCFT geometry into the bulk and reflected off the ETW-brane so as to return to the BCFT geometry (fig.~\ref{fig:Empty}).
The analog in the top-down duals are null geodesics launched from the $\ads_2$ conformal boundary of the $\ads_3\times S^3$ region in (\ref{eq:metric-AdS3S3}) into $\Sigma$, which, after moving through $\Sigma$, return back to the $\ads_3\times S^3$ region at $w=0$. Whether such geodesics exist was studied numerically in the 6d top-down solutions in \cite{Reeves:2021sab}. Here we give a simple analytic treatment and sharpen the conclusions.

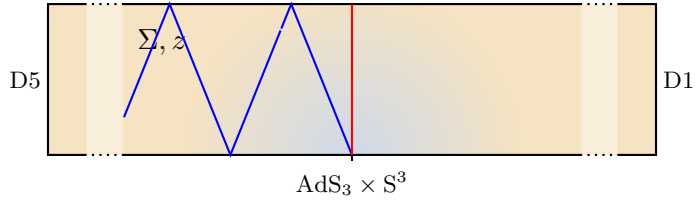
\begin{figure}
	\centering
	\begin{tikzpicture}[scale=1]
		\shade [left color=3dcolor!100,right color=3dcolor!100] (-3,0)  rectangle (3,-2);
		\begin{scope}
			\clip (-3,-2) rectangle (3,0);
			\fill[inner color=4dcolor!100,outer color=3dcolor,3dcolor] (0,-2) circle (2.2);	
		\end{scope}
		
		\draw[3dcolor,fill=3dcolor] (-3.5,0) rectangle (-4,-2);
		\draw[3dcolor,fill=3dcolor,opacity=0.6] (-3,0) rectangle (-3.5,-2);
		\draw[3dcolor,fill=3dcolor,opacity=0.6] (3,0) rectangle (3.5,-2);
		\draw[3dcolor,fill=3dcolor] (3.5,0) rectangle (4,-2);
		
		
		\draw[thick] (-3,0) -- (3,0);
		\draw[thick] (-3,-2) -- (3,-2);
		\draw[thick,dotted] (-3,0) -- (-3.5,0);
		\draw[thick,dotted] (-3,-2) -- (-3.5,-2);
		\draw[thick,dotted] (3,0) -- (3.5,0);
		\draw[thick,dotted] (3,-2) -- (3.5,-2);
		\draw[thick] (-3.5,0) -- (-4,0) -- (-4,-2) -- (-3.5,-2);
		\draw[thick] (3.5,0) -- (4,0) -- (4,-2) -- (3.5,-2);
		
		\node at (-2.5,-0.5) {$\Sigma, z$};
		\node at (4.3,-1) {\footnotesize D1};
		\node at (-4.3,-1) {\footnotesize D5};
		
		\draw[thick] (0,-1.92) -- (0,-2.08) node [anchor=north] {\footnotesize $\rm AdS_3\times S^3$};

		\draw[thick,blue] (0,-2) -- (-0.8,0) -- (-0.8-0.13,-2/0.8*0.13);
		\draw[thick,blue] (-0.8-0.14,-2/0.8*0.14) -- (-1.6,-2) -- (-2.4,0) --(-2.4-0.6,-2/0.8*0.6);
		\draw[thick,red] (0,-2)--(0,0);
	\end{tikzpicture}
	\caption{Geodesics \eqref{eq: geodesic-summary} starting at $z=0$ on $\Sigma$. When reaching one of the boundaries they are reflected and swapped to the antipodal point on the $S^2$. Generic geodesics do not return to the $\ads_3\times S^2$ region (blue), only the one along $\Re z=0$ does (red). \label{fig: strip}}
\end{figure}

The crux is that we can find a coordinate transformation on $\Sigma$ which makes $f_1^2/\rho^2$ constant. It maps the upper half plane parametrized by $w$ to a strip parametrized by a new coordinate $z$ defined as
\begin{align}\label{eq:poles-to-bndy-z}
	w&=\tanh z~,
	&
	\Sigma=\lbrace z\in\CC\,\vert\, 0\leq \Im(z)\leq \frac{\pi}{2}\rbrace~.
\end{align}
This maps the poles at $w=\pm 1$ to $\Re(z)=\pm\infty$ while the $\rm AdS_3\times S^3$ region at $w=0$ is mapped to $z=0$.
See fig.~\ref{fig: strip}.
The 6d metric becomes
\begin{align}\label{eq:geod-metric}
	ds^2&=f_1^2\left[ds^2_{\rm AdS_2}+\frac{H^2}{f_1^4} ds^2_{S^2}+\frac{\mu_0^2}{\kappa^2}|dz|^2\right]~,
\end{align}
where
\begin{align}\label{eqn:metric-z-new}
	f_1^4&=\frac{\kappa^2}{2}|\csch z|^4\left(\frac{4\kappa^2}{\mu_0^2}-\sinh^2(z-\bar z)\right)~,
	&
	H&=-\sqrt{2}\kappa \Im(\coth z)~.
\end{align}
In particular, since null geodesics are insensitive to the overall conformal factor, null geodesics without angular momentum on the $S^2$ experience an actual product geometry $\rm AdS_2\times\tilde\Sigma$, where $\tilde\Sigma$ is the strip parametrized by $z$ but with flat metric $ds^2_{\tilde\Sigma}=\mu_0^2\kappa^{-2}|dz|^2$.
As in the braneworld model, we take the $\ads_2$ part in (\ref{eq:geod-metric}) in global coordinates,
\begin{equation}
	ds_{\ads_2}^2=\frac{dv^2-dt^2}{\cos^2 v}~.
\end{equation}
This describes the dual BCFT on a hemisphere (see e.g.\ \cite{Uhlemann:2023oea} for a detailed discussion).

We now discuss geodesics without angular momentum on $S^2$ explicitly. 
Upon dropping a conformal factor in (\ref{eq:geod-metric}) the effective 4d geometry explored by null geodesics is
\begin{align}\label{eq:ds2eff-null}
	ds^2_{\rm eff}&=dv^2- dt^2+h(v)^2|dz|^2~, & 
	h(v)&=\frac{\mu_0}{\kappa}\cos v~.
\end{align}
The geodesic equations for curves parametrized by $z(\lambda)$, $u(\lambda)$, $t(\lambda)$ lead to
\begin{align}\label{eq:geod-0}
	t''&=0~, &
	v''-h(v)h'(v)|z'|^2&=0~,
	& 
	\frac{d}{d\lambda}\left(h(v)^2 z'(\lambda)\right)&=0~,
\end{align}
and the null condition is
\begin{align}
	v'^2+h(v)^2|z'|^2-t'^2&=0~.
\end{align}
The solutions for geodesics starting from $z=0$ are
\begin{align}\label{eq: geodesic-summary}
	\sin^2v&=\left(1-\frac{c_0^2\kappa^2}{\mu_0^2}\right)\sin^2(t-t_0)~
	&
	z&=	z_0+ \frac{\kappa}{\mu_0}e^{i\phi}\tan^{-1}\left(\frac{c_0\kappa}{\mu_0}\tan(t-t_0)\right) ~.
\end{align}
with integration constants $z_0$, $t_0$, $\phi$ and $c_0$ with $c_0^2<\frac{\mu_0^2}{\kappa^2}$.
These geodesics follow straight lines on $\Sigma$, as expected from the form of the geometry and illustrated in fig.~\ref{fig: strip}. 

Geodesics starting at $z=0$ follow straight lines to the boundary of $\Sigma$ at $\Im(z)=\frac{\pi}{2}$. This is not an actual boundary in the 10d geometry, as discussed above, since the $S^2$ collapses and the geometry closes off smoothly. The behavior of geodesics reaching $\Im(z)=\frac{\pi}{2}$ therefore follows from smoothness of the curves. Unlike in bottom-up models, it does not have to be imposed by hand. Smooth curves are reflected at $\Im(z)=\frac{\pi}{2}$ back into $\Sigma$, and reemerge on the antipodal point of $S^2$.\footnote{This is analogous to a straight line through the origin of $\RR^2$ in polar coordinates $(r,\varphi)$, which describe the flat $\RR^2$ geometry as warped product of an $S^1$ parametrized by $\varphi$ over $\RR_+$ parametrized by $r$: a straight line with constant $\varphi$ moves from $r=\infty$ towards $r=0$, where it is reflected to go back out to $r=\infty$ but on the antipodal point of the $S^1$ parametrized by $\varphi$.} They are smooth through the reflections as a result of how the motion on $\Sigma$ combines with the motion on the $S^2$ in the geometry (\ref{eq:geod-metric}). Generic geodesics bounce back and forth between the boundaries of $\Sigma$ towards one of the brane sources at $\Re(z)=\pm\infty$. This is shown as the blue curve in fig.~\ref{fig: strip}. Only the fine-tuned geodesics with $\phi=0$ in (\ref{eq: geodesic-summary}), shown as red line in fig.~\ref{fig: strip}, return to $z=0$.

To conclude this subsection and connect to its title, we compare the behavior of geodesics in the top-down construction to that in bottom-up braneworld models. We conclude the following: The strip geometry ($\Sigma$ with the $w\approx 0$ region removed, with $\ads_2$ and $S^2$ warped over it), which can be seen as microscopic version or `resolution' of the bottom-up ETW-brane, swallows almost all geodesics launched into it, rather than returning them indiscriminately. In this sense, `resolved ETW-branes are sticky'.

\subsection{Penrose limit for fine-tuned geodesics}

The only geodesics potentially leading to bulk-cone singularities are the ones moving on the strip $\Sigma$ along  $\Re z=0$ (the red line in fig.~\ref{fig: strip}). So we focus on these geodesics and derive the pp-wave emerging in the Penrose limit.

We introduce real coordinates on $\Sigma$, setting $z=x+i y$, and note that the geodesic equations for geodesics along $x=0$ can be written as
\bea \label{eq: geo-6d}
-t'^2+v'^2+\con\cos^2 v\, y'^2 &=0, ~&\text{null condition}\\
\con y' f_1^2 &=c_1 c_0, ~&\text{conserved $y$ momentum}\\
t'\frac{f_1^2}{\cos^2 v}&=c_1, ~&\text{conserved $t$ energy}
\eea
where $f_1$ is understood as $f_1(0,y)$. 
We set $c_1=1$ by rescaling the affine parameter.
A set of adapted coordinates $(U,V,W)$ can be defined as
\bea \label{eq:adapt-trans-6d}
dV & = dt + \dot{v}\,dv + c_0\,dy~,  \\
dU & = \frac{f_{1}^2\, \mu_0^2}{\kappa^2\, c_0}\,dy ~, \\
dW & = -\frac{c_0^2\, \kappa^2}{\mu_0^2\, \cos^2 v\, \dot{v}}\,dv + c_0\,dy~.
\eea
The null condition implies $\dot{v}^2=1-c_0^2\kappa^2\mu_0^{-2}\sec^2\! v$, which makes all three differential forms exact. So they can be integrated for $(U,V,W)$.

\begin{figure} 
	\centering
	\includegraphics[width=0.38\linewidth]{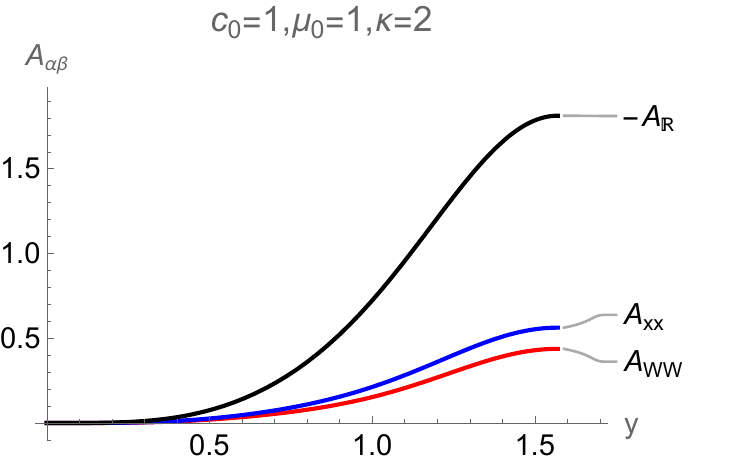}
	\qquad\qquad
	\includegraphics[width=0.38\linewidth]{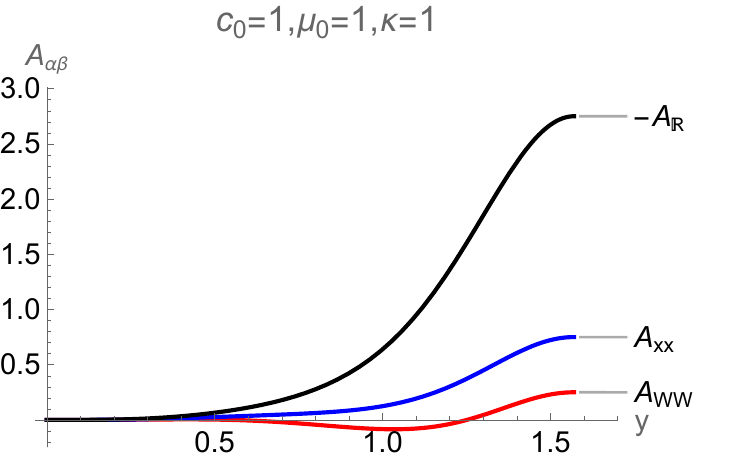}
	\caption{$A_{\alpha\beta}$ components as functions of $y$ with $\kappa=2$ (left) and $\kappa=1$ (right), and  $c_0=\mu_0=1$. The $A_{\alpha\beta}$ vanish at $y=0$ where the asymptotic $\ads_3\times S^3$ emerges, and are non-trivial otherwise. \label{fig:AWW-y-numeric}}
\end{figure}

Taking the Penrose limit for the 6d metric (\ref{eq:geod-metric}) amounts to rescaling the coordinates by a real parameter $a$,
\begin{equation}
	(U,V,W,x,ds^2_{S^2})\to(U,a^2 V, a W, ax, a^2 ds^2_{\RR^2} )~,
\end{equation}
and taking the limit $a\to 0$.
This leads to the pp-wave spacetime 
\begin{equation}
	ds^2=2dUdV+e_W^2(U)dW^2+e_x^2(U)dx^2+e_{\mrr}^2(U) ds^2_{\RR^2}~,
\end{equation}
where
\bea
	e_W^2(U)&=\left(\frac{\mu_0^2}{c_0^2\kappa^2}-\frac{1}{\cos^2 v}\right)f_1^2~,
	\qquad
	e_x^2(U)&=\frac{\mu_0^2}{\kappa^2}f_1^2~,
	\qquad
	e_{\mrr}^2 (U)&=\frac{H^2}{f_1^2}~.
\eea
Here $f_1$ and $H$ are understood as $f_1(0,y)$ and $H(0,y)$.
The coefficient of $f_1$ in  $e_W^2(U)$ is non-negative, since due to the null condition $1-c_0^2\kappa^2\mu_0^{-2}\sec^2\! v=\dot{v}^2\geq 0$.
In these expressions $v$ and $y$ are understood to be expressed as functions of $U$ through the geodesic equations. 
The pp-wave profile is, according to \eqref{eq: pp-wave-trans},
\begin{equation}
	A_{\alpha\alpha}=\frac{e_{\alpha}''(U)}{e_{\alpha}(U)}~.
\end{equation}
We evaluate this without solving explicitly for $v(U)$ and $y(U)$, by using
\begin{align}
	y'(U) &= \frac{c_0 \kappa^2}{\mu_0^2 f_1^2}~,
	&
	v'(U)& = 
	\frac{\cos^2\!v}{f_1^2}
	\sqrt{
		1
		- \frac{c_0^2\kappa^2}{\mu_0^2\cos^2 v}}~,
	&
\end{align}
which follow from the geodesic equations \eqref{eq: geo-6d}.
This leads to 
\bea\label{eq: Aww-D1-D5}
A_{WW}= 
\frac{2c_0^2\sin^4\! y}{\mu_0^2 \left( 4 \kappa^2 + \mu_0^2\sin^2(2y) \right)^3} \Big[
& 16 \kappa^6 + 8 \kappa^4 \mu_0^2(\cos(2y)-2\sin^2(2y)) 
\\
& 
+ \kappa^2 \mu_0^4 \sin^2(2y)(2\cos(2y)-9) - \mu_0^6 \sin^4(2y)
\Big] ~,\\
A_{xx}=
\frac{2 c_0^2 \kappa^2 \sin^4\!y}{\mu_0^2 (4 \kappa^2 + \mu_0^2 \sin^2(2y))^3} \Big[& 16 \kappa^4 + 8 \kappa^2 \mu_0^2 \cos(2y)\,( 2\cos(2y)+1)
\\
& 
	+ \mu_0^4\sin^2(2y)\,( 2\cos(2y)-1) \Big]~,\\
A_{\mrr} = \frac{2 c_0^2 \kappa^2 \sin^4\!y}{\mu_0^2 (4 \kappa^2 + \mu_0^2 \sin^2(2y))^3} \Big[& -48\kappa^4-8\kappa^2\mu_0^2 (2-\cos(2y)+2\cos^2(2y))
\\
& 
	 +\mu_0^4 \sin ^2(2y) (2 \cos (2 y)-1)]~.
\eea
Here $y$ is understood as $y(U)$, but we can also discuss the behavior of $A_{\alpha\alpha}$ in terms of $y$ itself.
All $A_{\alpha\alpha}$ approach $0$ in the $\ads_3\times S^3$ region at $y=0$, which is expected, but are non-trivial in the interior of $\Sigma$.
Plots for two sample parameter sets are shown in fig.~\ref{fig:AWW-y-numeric}.

\subsection{Large-$p_v$ suppression for fine-tuned geodesics}\label{sec:large-p-top-down}
We will not discuss the worldsheet computations for the profiles (\ref{eq: Aww-D1-D5}) in full detail. Instead, we recall that the main ingredient for demonstrating finiteness of the propagator in section~\ref{sec: resolve} was the large-$p_v$ behavior of the overlap between early and late vacua in (\ref{eq: overlap-vacuum}), which led to the bound (\ref{eq: uppder-bound}). So we focus on this large-$p_v$ limit.

We recall the equations of motion for the string fluctuations in the general pp-wave spacetime (\ref{eq:pp-wave-0}), in light-cone gauge $U=\alpha'p_v\tau$, given in (\ref{eq: string eom pp gen}), 
\begin{equation}\label{eq: string eom pp gen-rep}
	\left(\partial_\tau^2-\partial_\sigma^2\right) X^{\alpha}(\tau,\sigma)=\sum_{\beta} (\ap p_v)^2 A_{\alpha\beta}\left(\ap p_v \tau\right) X^{\beta}(\tau,\sigma)~,
\end{equation}
where the NS-NS two-form and dilaton were ignored.
This led for fluctuations in $W$ in the braneworld model where the pp-wave took a shockwave form to (\ref{eq:W-eom}),
\be\label{eq:W-eom-rep}
\lbra{\p_{\tau}^2-\p_{\sigma}^2}W(\tau,\sigma)-\ap p_v A \delta(\tau) W(\tau,\sigma)=0~.
\ee

We now show that the general pp-wave profile for the top-down D1/D5 BCFT dual in (\ref{eq: Aww-D1-D5}) reduces to this shockwave form in the limit  $p_v\rightarrow\infty$. We start with the components $A_{\alpha\alpha}$ given in (\ref{eq: Aww-D1-D5}) as function of $y$. The integrals
\begin{align}
	A^{\rm D1/D5}_{\alpha\alpha}&=2\int_0^{\pi/2}dy A_{\alpha\alpha}(y)
\end{align}
are absolutely convergent and can be evaluated explicitly,
\bea\label{eq:AD1D5}
	A^{\rm D1/D5}_{WW}&=\frac{\pi c_0^2}{8\kappa^2}a\left[a-\frac{20 a^4+81 a^2+58}{32 \left(a^2+1\right)^{3/2}} \right]~,\qquad
	a=\frac{2\kappa}{\mu_0}~,
	\\
	A^{\rm D1/D5}_{xx}&=\frac{3  \pi c_0^2 a	( 4a^4 + 5a^2+2)}{256 \,\kappa^2 \left(a^2+1\right)^{3/2}}~,\\
    A^{\rm D1/D5}_{\mrr}&=
    -\frac{\pi c_0^2  a \left(36 a^4+101 a^2+66\right) }{256 \kappa^2\left(a^2+1\right)^{3/2} }~.
\eea
For the string theory discussion the relevant function is $A_{\alpha\alpha}(y(U))$, and the statements about the integrals remain valid when changing variables to $y(U)$ and integrating over $U$. We note that $A^{\rm D1/D5}_{xx,\mrr}$ are both non-zero for non-zero $a$, while $A^{\rm D1/D5}_{WW}$ vanishes at $a\sim1.94$.

\begin{figure} 
	\centering
	\includegraphics[width=0.4\linewidth]{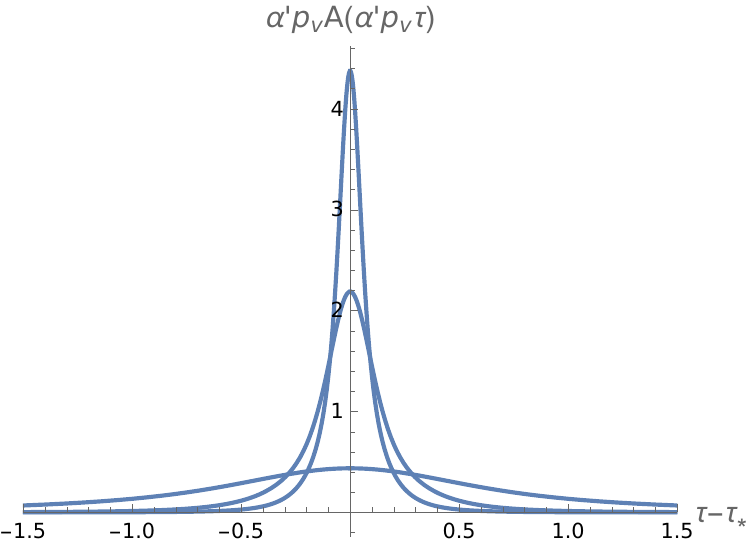}
	\caption{Plot of $\alpha'p_vA_{WW}(\alpha'p_v \tau)$ corresponding to the first case in fig.~\ref{fig:AWW-y-numeric}, for $p_v=1,5,10$. We solved for $U(y)$ using the second relation in (\ref{eq:adapt-trans-6d}), which can be expressed using elliptic integrals, and then used $U=\alpha'p_v\tau$ with $\alpha'=1$. The doubling of the domain is explained in the text.\label{fig: A-numeric}}
\end{figure}

When expressing $A_{\alpha\alpha}$ as function of $U$, we use an analogous doubling trick as in the braneworld models. That is, we extend the $y$ coordinate past $y=\frac{\pi}{2}$ and double the geometry. A geodesic turning around smoothly in the top-down geometry at $y=\frac{\pi}{2}$ then lifts to a geodesic that extends smoothly into the doubled geometry, proceeding to $y=\pi$ rather than back to $y=0$. The pp-wave profiles along the full geodesics are then $A_{\alpha\alpha}$, as e.g.\ in fig.~\ref{fig:AWW-y-numeric}, extended to the interval $y\in(0,\pi)$ by reflection across $y=\frac{\pi}{2}$.

With these preparations, we discuss the function $A_{\alpha\alpha}(y(U))$ with $U=\alpha' p_v \tau$, which is relevant for the worldsheet equations of motion. We note that the combinations
\begin{align}
	\alpha'p_v A_{\alpha\alpha}\left(\alpha'p_v \tau\right)
\end{align}
have finite integrals, namely $A^{\rm D1/D5}_{\alpha\alpha}$, for all $p_v$. They provide approximations to the identity when $p_v\rightarrow \infty$ (in the sense of identity for convolution). That is,
\begin{align}\label{eq:shock-emergence}
	\alpha'p_v A_{\alpha\alpha}\left(\alpha' p_v \tau\right)&\rightarrow A^{\rm D1/D5}_{\alpha\alpha}\delta(\tau)~,
\end{align}
with $A^{\rm D1/D5}_{\alpha\alpha}$ in (\ref{eq:AD1D5}).
This is illustrated in fig.~\ref{fig: A-numeric}. At least one of the $A^{\rm D1/D5}_{\alpha\alpha}$ is non-zero for all $\kappa$, $\mu_0$, and the equation for the corresponding transverse mode $X^{\alpha}$ in (\ref{eq: string eom pp gen-rep}) therefore reduces at large $p_v$ to the form (\ref{eq:W-eom-rep}) with $A\to A^{\rm D1/D5}_{\alpha\alpha}$. 

As a qualitative argument, we may use (\ref{eq:shock-emergence}) in the string action (\ref{eq: action-gf}) and the resulting equations of motion (\ref{eq: string eom pp gen-rep}). The discussion of section \ref{sec: resolve} would then lead to a bound on the two-point functions as in (\ref{eq: uppder-bound}) and resolve the bulk-cone singularities. A full string theory calculation would have to incorporate the remaining background fields in (\ref{eq: action-string-gen}), where e.g.\ the 10d 2-form $B_{\mu\nu}$ arises as a combination of 2-form fields in the 6d supergravity, and may also include the remaining spatial dimensions and worldsheet fermions. We will not do a full computation here and leave it at the more qualitative argument instead.

\smallskip

In closing we note two implications of the above discussion for bottom-up braneworld models. First, the discussion above shows that the shockwave pp-wave geometry underlying the braneworld analysis in sections \ref{sec: Pen-brane}, \ref{sec: resolve} emerges as a particular limit from the top-down construction. This provides some justification for the string theory computations in section~\ref{sec: resolve}. The second point concerns the generalized braneworld models in (\ref{eq:ds2-braneworld-gen}). The emergence of the shockwave form in the large-$p_v$ limit in (\ref{eq:shock-emergence}) did not rely on the detailed form of the pp-wave profiles $A_{\alpha\alpha}(U)$. It therefore also applies to the generalized braneworld models in (\ref{eq:ds2-braneworld-gen}), which also lead to more general pp-wave profiles $A_{\alpha\alpha}(U)$ compared to the shockwave in (\ref{eq: A-brane}). The large-$p_v$ behavior of the overlap between the early and late vacua therefore still takes the form in (\ref{eq: overlap-vacuum}) with a constant $A$. Since the large-$p_v$ behavior was the ingredient making the right hand side of (\ref{eq: uppder-bound}) finite, this leads to an upper bound on the stringy two-point function also in generalized braneworld models.

\section{Discussion}\label{sec: dis}

We discussed the resolution of bulk-cone singularities in braneworld models and top-down holographic duals for D1/D5 BCFTs. For the pp-wave spacetimes emerging in the Penrose limit associated with the bulk null geodesics that lead to bulk-cone singularities, we showed that they take a shockwave form in the simplest braneworld models based on undeformed AdS spacetimes, while in general braneworlds and in top-down constructions, string worldsheet modes with large momentum $p_v$ still experience pp-waves that effectively reduce to a shockwave form. The latter was the crucial ingredient in bounding the two-point functions.

\begin{figure}\label{fig:strip-comp}
	\centering
	
		\begin{tikzpicture}[scale=1]
			\shade [left color=3dcolor!80,right color=3dcolor!100] (-2,0)  rectangle (0.3,-2);
			\shade [ left color=3dcolor! 100, right color=4dcolor! 100] (0.3-0.01,0)  rectangle (2,-2);

			\draw[thick] (-2,0) -- (2,0);
			\draw[thick] (-2,-2) -- (2,-2);
			\draw[dashed] (2,-2) -- +(0,2);
			\draw (-2,-2) -- +(0,2);
			
			\node at (-1.5,-0.5) {$\Sigma, z$};
			\node at (2.5,-0.65) {\footnotesize $\rm AdS_3$};
			\node at (2.5,-1) {\footnotesize $\times$};
			\node at (2.5,-1.35) {\footnotesize $\rm S^3$};
			
			\draw[very thick] (0,-0.08) -- (0,0.08) node [anchor=south] {\footnotesize D1};
			\draw[thick] (0,-1.92) -- (0,-2.08) node [anchor=north] {\footnotesize D5};
			
		\end{tikzpicture}	
	\hskip 10mm
		\begin{tikzpicture}[scale=0.89]
			\node at (4,0){\includegraphics[width=0.28\linewidth]{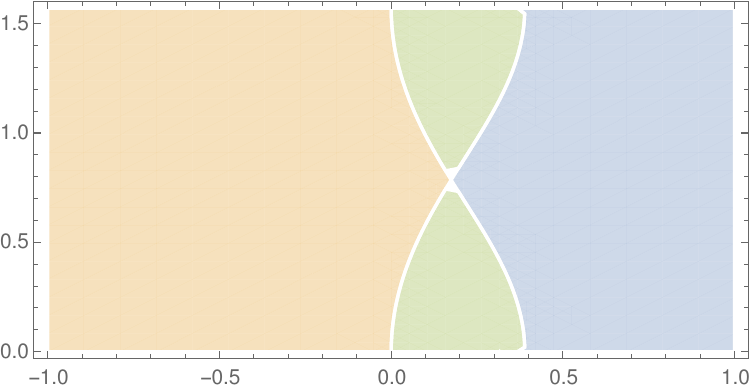}};
			\node at (5.8,0.1) {\small 4d};
			\node at (2.8,0.1) {\small 3d};
			\node at (2.4,0.8) {\small {\boldmath{$\Sigma$}}};

			\draw[very thick] (4+0.1,-1.1+0.1) -- (4+0.1,-1.1-0.1) node [anchor=north,yshift=0.5mm] {\small NS5};
			\draw[very thick] (4+0.1,1.25-0.1) -- (4+0.1,1.25+0.1) node [anchor=south,yshift=-0.5mm] {\small D5};
			
			\node at (6.9,0.4) {\footnotesize $\rm AdS_5$};
			\node at (6.9,0) {\footnotesize $\times$};
			\node at (6.9,-.4) {\footnotesize $\rm S^5$};
		\end{tikzpicture}
	
	\caption{Left: D1/D5 BCFT  dual of section \ref{eq:cap-sol} with coordinate transformation $w=e^{-z}$ so that $\Sigma=\lbrace z\in\CC\,\vert\, 0\leq \Im(z)\leq \pi\rbrace$. Right: $\Sigma=\lbrace z\in\CC\,\vert\, 0\leq \Im(z)\leq \frac{\pi}{2}\rbrace$ in the $\ads_4\times S^2\times S^2\times\Sigma$ holographic dual for a particular BCFT engineered by D3-branes ending on D5 and NS5 branes discussed in \cite{Uhlemann:2021nhu,Karch:2022rvr,Coccia:2021lpp}. The figure is from \cite{Coccia:2021lpp}, where, by studying the holographic representation of Wilson loops, an identification of a region of $\Sigma$ (shown in orange) with 3d observables was derived.}	
\end{figure}

We note that the top-down D1/D5 BCFT duals have similarities with holographic duals of 4d BCFTs based on $\mathcal N=4$ SYM with Gaiotto-Witten boundary conditions (though there are also qualitative differences). The duals for the 4d BCFTs have 10d geometries of the form $\ads_4\times S^2\times S^2\times \Sigma$ in Type IIB, and the similarities in their general form are highlighted in fig.~\ref{fig:strip-comp}. 
The figure also illustrates a result derived in \cite{Coccia:2021lpp}: the loose connection between some region of $\Sigma$ which has the asymptotic $\ads_{d+1}$ region removed and the bottom-up ETW brane can be made more quantitative by studying concrete observables. It would be interesting to establish a similar picture for the 2d BCFT duals, to make a more quantitative connection to the bottom-up models. More broadly, the field theory description of the 4d BCFTs is understood in great detail, and it would be desirable to understand the 2d BCFTs in more detail as well.

Other interesting things to explore about BCFT bulk-cone singularities include studying two-point functions and their singularities in field theory in connection with $1/\lambda$ effects, either in the 2d large-$c$ BCFTs or higher dimensional holographic BCFTs.
It may also be interesting to study finite coupling effects in the string theory computations.

\section*{Acknowledgments}	
DH thanks Andrea Conti, Marius Gerbershagen, and Juan Hernandez for helpful discussions.
DH is supported by FWO-Vlaanderen project G012222N and by the VUB Strategic Research
Program High-Energy Physics.

\appendix
\section{Penrose limit and adapted coordinates} \label{appendix: Penrose}
There is a well-defined limiting spacetime in the vicinity of null geodesics which generally takes the form of a pp-wave. This limit is called the Penrose limit \cite{Penrose:1976pw}.
The general way to take the Penrose limit for an arbitrary spacetime begins with finding an adapted set of coordinates for the null geodesic congruence so that the metric takes the form \cite{Blau:2024}
\begin{equation} \label{eq: pre-pp-standard}
	d s^2= d U d V+a\left(U, V, Y^{\gamma}\right) d V^2+2 b_{\alpha}\left(U, V, Y^{\gamma}\right) d V d Y^{\alpha}+g_{\alpha\beta}\left(U, V, Y^{\gamma}\right) d Y^{\alpha} d Y^{\beta}~,
\end{equation}
such that $\p_U$ is a null vector. $U,~V$ are null coordinates and $Y^{\alpha}$ are transverse coordinates.
Then the Penrose limit can be implemented by rescaling the coordinates and the metric,
\bea
(U,V,Y^{\alpha})&\to(U,a^2 V, a Y^{\alpha})~,\\
ds^2& \to a^2 ds^2~,\\
\eea
and taking the limit $a\to 0$.
Then the metric \eqref{eq: pre-pp-standard} becomes
\bea\label{eq: pp-wave-Rosen}
d s^2=2 d U d V +\bar{g}_{\alpha\beta}(U)d Y^{\alpha} d Y^{\beta}~.
\eea
These coordinates are called Rosen coordinates.
Then it is straightforward to do a further coordinate transformation to the Brinkham coordinates, in which the string action becomes quadratic with canonical kinetic terms
\bea \label{eq: pp-wave}
ds^2=2 dU dV+A_{\alpha\beta}(U)x^{\alpha} x^{\beta} dU^2+dx_{\alpha}^2~,
\eea
To find the adapted coordinate \cite{Patricot:2003dh,Blau:2024}, observe that in \eqref{eq: pre-pp-standard}
\bea
g_{UU}&=g_{U\alpha}=0~,\\
g_{UV}&=1~.
\eea
Then one can solve the equations to obtain the adapted coordinates,
\bea \label{eq: adapt-condition}
g^{\mu\nu}\p_{\mu}V\p_{\nu}V=&0~,\\
g^{\mu\nu}\p_{\mu}V\p_{\nu}Y^{\alpha}=&0~,\\
g^{\mu\nu}\p_{\mu}U\p_{\nu}V=&1~,
\eea
where $g_{\mn}$ is the metric component in the original coordinate and $(U,V,Y^{\alpha})$ are functions of the original coordinates, expressing the adapted coordinates in terms of the original coordinates.
Usually $V$ can be identified with the Jacobi function of the geodesics and $U$ is the affine parameter.

In this work we are concerned with special cases where the transverse components of the metric in Rosen coordinates, \eqref{eq: pp-wave-Rosen}, are diagonal,
\bea\label{eq: pp-wave-Rosen-dia}
ds^2=2dUdV+e_{\alpha}^2(U)\delta_{\alpha\beta} dY^{\alpha} dY^{\beta}~.
\eea
In this case the diagonal pp-wave profile in Brinkham coordinates can be obtained as
\bea\label{eq: pp-wave-trans}
A_{\alpha\beta}=\frac{e''_{\alpha}(U)}{e_{\alpha} (U)}\delta_{\alpha\beta}~,
\eea
where the indices are not summed.
The pp-wave profile $A_{\alpha\beta}$ can be also read off directly from the components of the Riemann tensor in the adapted coordinate \cite{Blau:2024}.

\bibliographystyle{JHEP}
\bibliography{BCFT.bib}

\providecommand{\href}[2]{#2}\begingroup\raggedright\begin{thebibliography}{10}

\bibitem{Maldacena:2015iua}
J.~Maldacena, D.~Simmons-Duffin, and A.~Zhiboedov, {\it {Looking for a bulk
  point}},  {\em JHEP} {\bf 01} (2017) 013,
  [\href{https://arxiv.org/pdf/1509.03612}{{\tt arXiv:1509.03612}}].

\bibitem{Gary:2009ae}
M.~Gary, S.~B. Giddings, and J.~Penedones, {\it {Local bulk S-matrix elements
  and CFT singularities}},  {\em Phys. Rev. D} {\bf 80} (2009) 085005,
  [\href{https://arxiv.org/pdf/0903.4437}{{\tt arXiv:0903.4437}}].

\bibitem{Hubeny:2006yu}
V.~E. Hubeny, H.~Liu, and M.~Rangamani, {\it {Bulk-cone singularities \&
  signatures of horizon formation in AdS/CFT}},  {\em JHEP} {\bf 01} (2007)
  009, [\href{https://arxiv.org/pdfhep-th/0610041}{{\tt hep-th/0610041}}].

\bibitem{Dodelson:2023nnr}
M.~Dodelson, C.~Iossa, R.~Karlsson, A.~Lupsasca, and A.~Zhiboedov, {\it {Black
  hole bulk-cone singularities}},  {\em JHEP} {\bf 07} (2024) 046,
  [\href{https://arxiv.org/pdf/2310.15236}{{\tt arXiv:2310.15236}}].

\bibitem{Dodelson:2020lal}
M.~Dodelson and H.~Ooguri, {\it {Singularities of thermal correlators at strong
  coupling}},  {\em Phys. Rev. D} {\bf 103} (2021), no.~6 066018,
  [\href{https://arxiv.org/pdf/2010.09734}{{\tt arXiv:2010.09734}}].

\bibitem{Karch:2000ct}
A.~Karch and L.~Randall, {\it {Locally localized gravity}},  {\em JHEP} {\bf
  05} (2001) 008, [\href{https://arxiv.org/pdfhep-th/0011156}{{\tt
  hep-th/0011156}}].

\bibitem{Karch:2000gx}
A.~Karch and L.~Randall, {\it {Open and closed string interpretation of SUSY
  CFT's on branes with boundaries}},  {\em JHEP} {\bf 06} (2001) 063,
  [\href{https://arxiv.org/pdfhep-th/0105132}{{\tt hep-th/0105132}}].

\bibitem{Takayanagi:2011zk}
T.~Takayanagi, {\it {Holographic Dual of BCFT}},  {\em Phys. Rev. Lett.} {\bf
  107} (2011) 101602, [\href{https://arxiv.org/pdf/1105.5165}{{\tt
  arXiv:1105.5165}}].

\bibitem{Fujita:2011fp}
M.~Fujita, T.~Takayanagi, and E.~Tonni, {\it {Aspects of AdS/BCFT}},  {\em
  JHEP} {\bf 11} (2011) 043, [\href{https://arxiv.org/pdf/1108.5152}{{\tt
  arXiv:1108.5152}}].

\bibitem{Reeves:2021sab}
W.~Reeves, M.~Rozali, P.~Simidzija, J.~Sully, C.~Waddell, and D.~Wakeham, {\it
  {Looking for (and not finding) a bulk brane}},  {\em JHEP} {\bf 12} (2021)
  002, [\href{https://arxiv.org/pdf/2108.10345}{{\tt arXiv:2108.10345}}].

\bibitem{Kastikainen:2021ybu}
J.~Kastikainen and S.~Shashi, {\it {Structure of holographic BCFT correlators
  from geodesics}},  {\em Phys. Rev. D} {\bf 105} (2022), no.~4 046007,
  [\href{https://arxiv.org/pdf/2109.00079}{{\tt arXiv:2109.00079}}].

\bibitem{Gaiotto:2008sa}
D.~Gaiotto and E.~Witten, {\it {Supersymmetric Boundary Conditions in N=4 Super
  Yang-Mills Theory}},  {\em J. Statist. Phys.} {\bf 135} (2009) 789--855,
  [\href{https://arxiv.org/pdf/0804.2902}{{\tt arXiv:0804.2902}}].

\bibitem{Gaiotto:2008ak}
D.~Gaiotto and E.~Witten, {\it {S-Duality of Boundary Conditions In N=4 Super
  Yang-Mills Theory}},  {\em Adv. Theor. Math. Phys.} {\bf 13} (2009), no.~3
  721--896, [\href{https://arxiv.org/pdf/0807.3720}{{\tt arXiv:0807.3720}}].

\bibitem{DHoker:2007zhm}
E.~D'Hoker, J.~Estes, and M.~Gutperle, {\it {Exact half-BPS Type IIB interface
  solutions. I. Local solution and supersymmetric Janus}},  {\em JHEP} {\bf 06}
  (2007) 021, [\href{https://arxiv.org/pdf/0705.0022}{{\tt arXiv:0705.0022}}].

\bibitem{DHoker:2007hhe}
E.~D'Hoker, J.~Estes, and M.~Gutperle, {\it {Exact half-BPS Type IIB interface
  solutions. II. Flux solutions and multi-Janus}},  {\em JHEP} {\bf 06} (2007)
  022, [\href{https://arxiv.org/pdf/0705.0024}{{\tt arXiv:0705.0024}}].

\bibitem{Aharony:2011yc}
O.~Aharony, L.~Berdichevsky, M.~Berkooz, and I.~Shamir, {\it {Near-horizon
  solutions for D3-branes ending on 5-branes}},  {\em Phys. Rev. D} {\bf 84}
  (2011) 126003, [\href{https://arxiv.org/pdf/1106.1870}{{\tt
  arXiv:1106.1870}}].

\bibitem{Assel:2011xz}
B.~Assel, C.~Bachas, J.~Estes, and J.~Gomis, {\it {Holographic Duals of D=3 N=4
  Superconformal Field Theories}},  {\em JHEP} {\bf 08} (2011) 087,
  [\href{https://arxiv.org/pdf/1106.4253}{{\tt arXiv:1106.4253}}].

\bibitem{He:2024djr}
D.~He and C.~F. Uhlemann, {\it {Solving $ \mathcal{N} $ = 4 SYM BCFT matrix
  models at large N}},  {\em JHEP} {\bf 12} (2024) 164,
  [\href{https://arxiv.org/pdf/2409.13016}{{\tt arXiv:2409.13016}}].

\bibitem{He:2025etu}
D.~He and C.~F. Uhlemann, {\it {One-point functions for doubly-holographic
  BCFTs and backreacting defects}},
  \href{https://arxiv.org/pdf/2501.07630}{{\tt arXiv:2501.07630}}.

\bibitem{Chiodaroli:2011nr}
M.~Chiodaroli, E.~D'Hoker, Y.~Guo, and M.~Gutperle, {\it {Exact half-BPS
  string-junction solutions in six-dimensional supergravity}},  {\em JHEP} {\bf
  12} (2011) 086, [\href{https://arxiv.org/pdf/1107.1722}{{\tt
  arXiv:1107.1722}}].

\bibitem{Chiodaroli:2011fn}
M.~Chiodaroli, E.~D'Hoker, and M.~Gutperle, {\it {Simple Holographic Duals to
  Boundary CFTs}},  {\em JHEP} {\bf 02} (2012) 005,
  [\href{https://arxiv.org/pdf/1111.6912}{{\tt arXiv:1111.6912}}].

\bibitem{Chiodaroli:2012vc}
M.~Chiodaroli, E.~D'Hoker, and M.~Gutperle, {\it {Holographic duals of Boundary
  CFTs}},  {\em JHEP} {\bf 07} (2012) 177,
  [\href{https://arxiv.org/pdf/1205.5303}{{\tt arXiv:1205.5303}}].

\bibitem{Karch:2022rvr}
A.~Karch, H.~Sun, and C.~F. Uhlemann, {\it {Double holography in string
  theory}},  {\em JHEP} {\bf 10} (2022) 012,
  [\href{https://arxiv.org/pdf/2206.11292}{{\tt arXiv:2206.11292}}].

\bibitem{Penrose:1976pw}
R.~Penrose, {\it {Any Space-Time has a Plane Wave as a Limit}},  in {\em
  Differential Geometry and Relativity} (M.~Cahen and M.~Flato, eds.),
  pp.~271--275.
\newblock Reidel, Dordrecht, 1976.

\bibitem{Metsaev:2001bj}
R.~R. Metsaev, {\it {Type IIB Green-Schwarz superstring in plane wave
  Ramond-Ramond background}},  {\em Nucl. Phys. B} {\bf 625} (2002) 70--96,
  [\href{https://arxiv.org/pdfhep-th/0112044}{{\tt hep-th/0112044}}].

\bibitem{Blau:2024}
M.~Blau, {\it Plane waves and penrose limits},  2024.
\newblock Lecture notes, 89 pages.

\bibitem{Berenstein:2002jq}
D.~E. Berenstein, J.~M. Maldacena, and H.~S. Nastase, {\it {Strings in flat
  space and pp waves from N=4 superYang-Mills}},  {\em JHEP} {\bf 04} (2002)
  013, [\href{https://arxiv.org/pdfhep-th/0202021}{{\tt hep-th/0202021}}].

\bibitem{Chaney:2024bgx}
A.~Chaney and C.~F. Uhlemann, {\it {BMN-like sectors in 4d $\mathcal N=4$ SYM
  with boundaries and interfaces}},
  \href{https://arxiv.org/pdf/2408.12651}{{\tt arXiv:2408.12651}}.

\bibitem{Martinec:2020cml}
E.~J. Martinec and N.~P. Warner, {\it {The Harder They Fall, the Bigger They
  Become: Tidal Trapping of Strings by Microstate Geometries}},  {\em JHEP}
  {\bf 04} (2021) 259, [\href{https://arxiv.org/pdf/2009.07847}{{\tt
  arXiv:2009.07847}}].

\bibitem{Siopsis:2002vw}
G.~Siopsis, {\it {Holography in the Penrose limit of AdS space}},  {\em Phys.
  Lett. B} {\bf 545} (2002) 169--174,
  [\href{https://arxiv.org/pdfhep-th/0205302}{{\tt hep-th/0205302}}].

\bibitem{Mazac:2018biw}
D.~Maz\'a\v{c}, L.~Rastelli, and X.~Zhou, {\it {An analytic approach to
  BCFT$_{d}$}},  {\em JHEP} {\bf 12} (2019) 004,
  [\href{https://arxiv.org/pdf/1812.09314}{{\tt arXiv:1812.09314}}].

\bibitem{Omiya:2021olc}
H.~Omiya and Z.~Wei, {\it {Causal structures and nonlocality in double
  holography}},  {\em JHEP} {\bf 07} (2022) 128,
  [\href{https://arxiv.org/pdf/2107.01219}{{\tt arXiv:2107.01219}}].

\bibitem{deVega:1988ts}
H.~J. de~Vega and N.~G. Sanchez, {\it {Quantum String Scattering in the
  Aichelburg-sexl Geometry}},  {\em Nucl. Phys. B} {\bf 317} (1989) 706--730.

\bibitem{Giddings:2007bw}
S.~B. Giddings, D.~J. Gross, and A.~Maharana, {\it {Gravitational effects in
  ultrahigh-energy string scattering}},  {\em Phys. Rev. D} {\bf 77} (2008)
  046001, [\href{https://arxiv.org/pdf/0705.1816}{{\tt arXiv:0705.1816}}].

\bibitem{Horowitz:1990sr}
G.~T. Horowitz and A.~R. Steif, {\it {Strings in strong gravitational fields}},
   {\em Phys. Rev. D} {\bf 42} (1990) 1950--1959.

\bibitem{DeWitt:1975ys}
B.~S. DeWitt, {\it {Quantum Field Theory in Curved Space-Time}},  {\em Phys.
  Rept.} {\bf 19} (1975) 295--357.

\bibitem{Cohen:1985sm}
A.~G. Cohen, G.~W. Moore, P.~C. Nelson, and J.~Polchinski, {\it {An Off-Shell
  Propagator for String Theory}},  {\em Nucl. Phys. B} {\bf 267} (1986)
  143--157.

\bibitem{Green:1987mn}
M.~B. Green, J.~H. Schwarz, and E.~Witten, {\em {Superstring Theory. Vol. 2:
  Loop Amplitudes, Anomalies and Phenomenology}}.
\newblock Cambridge University Press, 1987.

\bibitem{Coccia:2021lpp}
L.~Coccia and C.~F. Uhlemann, {\it {Mapping out the internal space in AdS/BCFT
  with Wilson loops}},  {\em JHEP} {\bf 03} (2022) 127,
  [\href{https://arxiv.org/pdf/2112.14648}{{\tt arXiv:2112.14648}}].

\bibitem{Uhlemann:2023oea}
C.~F. Uhlemann and M.~Wang, {\it {Splitting interfaces in 4d $ \mathcal{N} $ =
  4 SYM}},  {\em JHEP} {\bf 12} (2023) 053,
  [\href{https://arxiv.org/pdf/2307.08740}{{\tt arXiv:2307.08740}}].

\bibitem{Uhlemann:2021nhu}
C.~F. Uhlemann, {\it {Islands and Page curves in 4d from Type IIB}},  {\em
  JHEP} {\bf 08} (2021) 104, [\href{https://arxiv.org/pdf/2105.00008}{{\tt
  arXiv:2105.00008}}].

\bibitem{Patricot:2003dh}
C.~Patricot, {\it {Kaigorodov spaces and their Penrose limits}},  {\em Class.
  Quant. Grav.} {\bf 20} (2003) 2087--2102,
  [\href{https://arxiv.org/pdfhep-th/0302073}{{\tt hep-th/0302073}}].

\end{thebibliography}\endgroup

\end{document}